\documentclass[manuscript]{aastex}

\usepackage{amsmath,amssymb}
\usepackage{array,multirow}
\usepackage{hyperref,url}
\usepackage{afterpage}
\usepackage{pdflscape}
\begin{document}

\title{Planet Occurrence Rate Density Models Including Stellar Effective Temperature}

\author{Daniel Garrett}
\affil{Sibley School of Mechanical and Aerospace Engineering, Cornell University, Ithaca, NY 14853, USA}
\affil{Carl Sagan Institute, Cornell University, Ithaca, NY 14853, USA}
\email{dg622@cornell.edu}

\author{Dmitry Savransky}
\affil{Sibley School of Mechanical and Aerospace Engineering, Cornell University, Ithaca, NY 14853, USA}
\affil{Carl Sagan Institute, Cornell University, Ithaca, NY 14853, USA}
\email{ds264@cornell.edu}

\author{Rus Belikov}
\affil{NASA Ames Research Center, USA}
\email{ruslan.belikov-1@nasa.gov}

\begin{abstract}
    We present planet occurrence rate density models fit to Kepler data as a function of semi-major axis, planetary radius, and stellar effective temperature. We find that occurrence rates for M type stars with lower effective temperature do not follow the same trend as F, G, and K type stars when including a polynomial function of effective temperature in an occurrence rate density model and a better model fit includes a break in effective temperature. Our model fit for M type stars consists of power laws on semi-major axis and planetary radius. Our model fit for F, G, and K type stars consists of power laws on semi-major axis and planetary radius broken at 2.771$ R_\oplus $ and a quadratic function of stellar effective temperature. Our models show agreement with published occurrence rate studies and are the first to explicitly include stellar effective temperature as a variable. By introducing stellar effective temperature into our occurrence rate density models, we enable more accurate occurrence rate predictions for individual stars in mission simulation and science yield calculations for future and proposed exoplanet finding missions.
\end{abstract}


\section{Introduction}\label{sec:intro}
Exoplanet-finding surveys have shown that planets are common \citep{winn2015occurrence,winn2018planet}. From these surveys the occurrence rate of planets may be inferred by applying a completeness or detection efficiency correction (correcting for observational or algorithmic biases which led to missed detections of real planets, e.g. \cite{christiansen2015measuring}) to the number of planets discovered with properties binned in a specified range (typically mass and orbital period for radial velocity, as in \cite{cumming2008keck}, or planet radius and period for transit surveys, as in \cite{howard2012planet}). Accurately determining occurrence rates requires a large sample of target stars, planet detections, and knowledge of specific instrument effects \citep{tabachnik2002maximum,youdin2011exoplanet,batalha2014exploring}. A model fit to occurrence rate densities can then be used to interpolate or extrapolate occurrence rates for a desired range of parameters \citep{catanzarite2011occurrence,traub2011terrestrial,biller2013gemini,kopparapu2013habitable,petigura2013prevalence,burke2015terrestrial}. For direct imaging surveys, such as the proposed Wide-Field Infrared Survey Telescope's (WFIRST) Coronagraph Instrument \citep{spergel2015wide}, these occurrence rate densities can be used to generate planet populations for completeness studies \citep{brown2005single,brown2010new,garrett2016analytical}, exoplanet yield maximization \citep{stark2014maximizing}, or full mission simulations \citep{savransky2015wfirst}.

The earliest models for planet occurrence rate densities were based on radial velocity surveys. \cite{tabachnik2002maximum} fit a joint power-law model to the occurrence rate densities of the 72 planets detected up to that point via the radial velocity method with $ m_p < 10 M_J $ and $ P > 2 $ days, where $ m_p $ is the mass of the planet and $ P $ is the orbital period. \cite{cumming2008keck} established a benchmark model for occurrence rates by fitting a joint power law to occurrence rate densities for $ m_p > 0.3 M_J $ and $ P < 2000 $ days from a radial velocity survey consisting of 600 F, G, K, and M type stars monitored over eight years. \cite{howard2010occurrence} extended the results of \cite{cumming2008keck} to lower masses by fitting a joint power law to the occurrence rate densities of planets for 166 G and K type stars with $ 3 M_\oplus < m_p < 1000 M_\oplus $ and $ P < 50 $ days. \cite{bryan2016statistics} fit a power law on mass, $ 0.05 M_J < m_p < 1000 M_J $, and semi-major axis, $ 1 AU < a < 500 AU $, to data collected at the Keck Observatory as a part of the California Planet Study \citep{howard2010california}.

The Kepler mission \citep{batalha2014exploring,borucki2017kepler} has yielded thousands of planet discoveries via the transit method and enabled investigation of occurrence rates of Earth-like planets. Many studies have fit occurrence rate density models to various period and radius ranges, stellar types, and Kepler data releases. \cite{youdin2011exoplanet}, \cite{burke2015terrestrial}, and \cite{mulders2018exoplanet} fit occurrence rate densities on period and planetary radius, $ R_p $, jointly. A number of studies fit occurrence rate densities only dependent on period \citep{howard2012planet,catanzarite2011occurrence,traub2011terrestrial,dong2013fast,petigura2013plateau,silburt2015statistical} or planetary radius \citep{howard2012planet,catanzarite2011occurrence,morton2014radius}. Primarily, these models have been power laws, however, \cite{berta2013constraints} fit a joint model which has power law and sigmoid terms and \cite{morton2014radius} used a weighted kernel density estimation (wKDE) approach. A summary of these models based on period and planetary radius along with the fit type, period and planetary radius ranges, stellar spectral type, and Kepler release data is given in Table \ref{tab:literature}. Models based on Kepler data are not restricted to period and planetary radius. Recently, \cite{pascucci2018universal} fit broken power laws to occurrence rate density dependent on planet-to-star mass ratio for F, G, K, and M type stars using Kepler Q1-Q17 data.

\afterpage{
\begin{landscape}
\begin{table}
    \small
    \caption{Occurrence rate density model fits on period and planetary radius from the Kepler literature.}
    \centering
    \begin{tabular}{cccccc}
        \hline\hline
        Source & Fit Type & $ P $ (days) & $ R_p $ $ \left(R_\oplus\right) $ & Spectral Type & Kepler Data \\
        \hline
        \cite{youdin2011exoplanet} & Joint power-law & $ \left[0.5,50\right] $ & $ \left[0.5,20\right] $ & G, K & Q0--Q2 \\
        \hline
        \multirow{2}{*}{\cite{howard2012planet}} & Power-law on $ P $  & $ \left[0.68,50\right] $  & $ \left[2,32\right] $ & G, K & Q0--Q2 \\
        & Power-law on $ R_p $ & $ \left[0.68,50\right] $  & $ \left[2,32\right] $ & G, K & Q0--Q2 \\
        \hline
        \multirow{2}{*}{\cite{catanzarite2011occurrence}} & Power-law on $ P $ & $ \left[0.68,132\right] $ & $ \left[2,4\right] $ & F, G, K & Q0--Q5 \\
        & Power-law on $ R_p $ & $ \left[0.68,132\right] $ & $ \left[2,4\right] $ & F, G, K & Q0--Q5 \\
        \hline
        \cite{traub2011terrestrial} & Power-law on $ P $ & $ \left[3,42\right] $ & $ \left[0.6,40\right] $ & F, G, K & Q0--Q5 \\
        \hline
        \cite{berta2013constraints} & Joint sigmoid & $ \left[0.25,50\right] $ & $ \left[0.8,4.5\right] $ & M & Q1--Q6 \\
        \hline
        \multirow{4}{*}{\cite{dong2013fast}} & Power-law on $ P $ & $ \left[10,250\right] $ & $ \left[1,2\right] $ & F, G, K & Q1-Q6 \\
        & Power-law on $ P $ & $ \left[10,250\right] $ & $ \left[2,4\right] $ & F, G, K & Q1-Q6 \\
        & Power-law on $ P $ & $ \left[10,250\right] $ & $ \left[4,8\right] $ & F, G, K & Q1-Q6 \\
        & Power-law on $ P $ & $ \left[10,250\right] $ & $ \left[8,16\right] $ & F, G, K & Q1-Q6 \\
        \hline
        \multirow{2}{*}{\cite{petigura2013plateau}} & Power-law on $ P $ & $ \left[5,10.8\right] $ & $ \left[1,8\right] $ & G, K & Q1--Q9 \\
        & Power-law on $ P $ & $ \left[10.8,50\right] $ & $ \left[1,8\right] $ & G, K & Q1--Q9 \\
        \hline
        \cite{morton2014radius} & wKDE for $ R_p $ & $ \left[0.68,150\right] $ & $ \left[0.3,4\right] $ & M & Q1--Q12 \\
        \hline
        \cite{burke2015terrestrial} & Joint power-law & $ \left[50,300\right] $ & $ \left[0.75,2.5\right] $ & G, K & Q1--Q16 \\
        \hline
        \cite{silburt2015statistical} & Power-law on $ P $ & $ \left[20,200\right] $ & $ \left[1,4\right] $ & F, G, K & Q1--Q16 \\
        \hline
        \cite{mulders2018exoplanet} & Joint power-law & $ \left[2,400\right] $ & $ \left[0.5,6\right] $ & F, G, K, M & Q1--Q17 \\
        \hline
    \end{tabular}
    \label{tab:literature}
\end{table}
\end{landscape}
}

Direct imaging surveys are sensitive to large planets at large separations from their host stars \citep{bowler2018occurrence} and complement other exoplanet detection techniques. Although fewer detections have been made via direct imaging, it has been shown that occurrence rate models from radial velocity or transit surveys cannot be extrapolated to the region probed by direct imaging and constraints on the occurrence rates at wide separations have been determined \citep{lafreniere2007gemini,nielsen2008constraints,nielsen2010uniform,biller2013gemini,wahhaj2013gemini,brandt2014statistical,vigan2017vlt}. These constraints have been used to synthesize models from microlensing, radial velocity, and direct imaging surveys \citep{clanton2014synthesizing,clanton2016synthesizing}.

All of the occurrence rate models mentioned are only dependent on planet and orbital properties. However, occurrence rates are dependent on stellar properties like metallicity and effective temperature \citep{winn2018planet,mulders2018planet}. \cite{marcy2005observed} showed that planet occurrence rises with stellar metalicity. Radial velocity surveys have shown that giant planet occurrence rates are associated with higher stellar metallicity \citep{gonzalez1997stellar,santos2004spectroscopic,valenti2005spectroscopic,mayor2011harps,reffert2015precise} while smaller planets are not \citep{sousa2008spectroscopic,buchhave2012abundance,beauge2012emerging}. \cite{schlaufman2018evidence} found this association with high metallicity to be weaker. For periods shorter than 10 days, small planets are associated with higher metallicities \citep{mulders2016super,petigura2018california,wilson2018elemental}. \cite{zhu2016dependence} found that the planet-metallicity correlation for small planets could be the same as that for giant planets. Using Kepler data, \cite{petigura2018california} fit an occurrence rate density model to period and stellar metallicity. 

Kepler data has been used to investigate the effect of stellar effective temperature, $ T_{eff} $, on planet occurrence rates. \cite{traub2011terrestrial} found that for periods less than 42 days, the occurrence rates for terrestrial planets are roughly the same, ice giants vary by a factor of two, and gas giants rapidly drop with $ T_{eff} $ for F, G, and K type stars. \cite{howard2012planet} and  \cite{mulders2015stellar,mulders2015increase} found that smaller planets occur more frequently around stars with lower $ T_{eff} $. \cite{dressing2013occurrence,dressing2015occurrence} supported this finding by investigating the occurrence rates of small planets around M dwarfs. \cite{fressin2013false} investigated the effect of stellar mass, which is related to $ T_{eff} $, on the occurrence rate of planets in various radius bins, and found that giant planet (6--22$ R_\oplus $) occurrence rates increase with $ T_{eff} $ for M, K, and G type stars and then decrease with $ T_{eff} $ for F type stars.  The same work found no dependence on $ T_{eff} $ for large Neptunes (4--6$ R_\oplus $) and small Neptunes (2--4$ R_\oplus $).

We present planet occurrence rate density models fit to previous occurrence rate calculations from the Kepler literature and data from NASA's Exoplanet Program Analysis Group (ExoPAG) Science Analysis Group 13\footnote{\url{https://exoplanets.nasa.gov/exep/exopag/sag/}} (hereafter SAG13) which include the planet parameters of semi-major axis, $ a $, and planetary radius, $ R_p $, while also incorporating $ T_{eff} $ (Section \ref{sec:models}). We discuss our data selection (Section \ref{sec:data}), model fitting process (Section \ref{sec:fits}), and results (Section \ref{sec:results}). Finally, we compare our models to data from the literature (Section \ref{sec:discussion}).

\section{Planet Occurrence Rate Models with Stellar Effective Temperature}\label{sec:models}
\cite{mulders2015stellar} showed that by stretching the semi-major axis and multiplying the overall occurrence rate by a fractional value dependent on $ T_{eff} $, the occurrence rates of 1--4$ R_\oplus $ planets for M, K, and G type stars collapse onto the occurrence rates of F stars (see \cite{mulders2015stellar} Figure 4). The scaling relationship between semi-major axis and stellar mass for the cutoff of occurrence rates near a period of 10 days for different spectral type stars is given by $ a \propto M^{1/3} $ \citep{mulders2015stellar,lee2017magnetospheric}. Since stellar mass can be represented as a function of $ T_{eff} $, we represent both the semi-major axis stretching and overall occurrence rate scaling as a single function of $ T_{eff} $. We wish to reference our model to solar and Earth values, so we normalize by $ T_{eff,\odot} = 5772 K $, $ a_\oplus = 1 AU $, and $ R_\oplus $. We define 
\begin{equation}
    t \triangleq \frac{T_{eff}}{T_{eff,\odot}} - 1
\end{equation}
to work with numerical values on a similar scale. We define the $ T_{eff} $ dependent portion of our occurrence rate density model as a power series with as many terms as necessary in $ t $ 
\begin{equation}
    g\left(t\right) = 1 + \gamma t + \mu t^2 + \nu t^3 + \ldots
\end{equation}
Our simple occurrence rate density model is given as
\begin{equation}
    \frac{\partial^2 \eta}{\partial \ln a \partial \ln R_p} = C\left(\frac{a}{a_\oplus}\right)^\alpha\left(\frac{R_p}{R_\oplus}\right)^\beta g\left(t\right) 
\end{equation}
where $ C $, $ \alpha $, $ \beta $, as well as the values and number of terms in $ g\left(t\right) $ are determined by fitting the model to data. In addition to this simple model, we investigate a break radius model similar to \cite{burke2015terrestrial}
\begin{equation}
    \frac{\partial^2 \eta}{\partial \ln a \partial \ln R_p} = \begin{cases}
        C_0\left(\frac{a}{a_\oplus}\right)^{\alpha_0}\left(\frac{R_p}{R_\oplus}\right)^{\beta_0} g_0\left(t\right) & R_p < R_b \\
        C_1\left(\frac{a}{a_\oplus}\right)^{\alpha_1}\left(\frac{R_p}{R_\oplus}\right)^{\beta_1} g_1\left(t\right) & R_p \geq R_b
    \end{cases}
\end{equation}
where the break radius, $ R_b $, as well as two sets of constants for $ g_0\left(t\right) $ and $ g_1\left(t\right) $ are determined via fitting to data.

\section{Data Selection}\label{sec:data}
We seek occurrence rate data binned on a wide range of planet properties (orbital period and planetary radius) and $ T_{eff} $. The SAG13 effort collected tables of occurrence rates from Kepler data and defined a standard grid of planetary radius, orbital period, and stellar type. The occurrence rates came from data and models from peer-reviewed publications \citep{petigura2013prevalence,foreman2014exoplanet,burke2015terrestrial,dressing2015occurrence,mulders2015increase,mulders2015stellar,mulders2016super,traub2016kepler,fulton2017california} and unpublished tables generated by Natalie Batalha, Ruslan Belikov, Joseph Catanzarite, Will Farr, and Ravi Kopparapu using the Q1-Q16 or Q1-Q17 planet candidates, DR24 star properties catalog, and Kepler completeness curves released in the fall of 2015. 

The SAG13 standard planetary radius-period grid consists of uniformly spaced bins in log radius and period, where the $ i $th planet radius bin is given by
\begin{equation}
    R_{p_i} = \left[1.5^{i-2},1.5^{i-1}\right)R_\oplus
\end{equation}
(bin edges $ \left[0.67,1.0,1.5,2.3,3.4,...\right]R_\oplus $) and the $ j $th orbital period bin given by
\begin{equation}
    P_j = 10\times\left[2^{j-1},2^j\right) \mathrm{days}
\end{equation}
(bin edges $ \left[10,20,40,80,160,...\right]$days). Unless otherwise indicated, the stellar types are grouped as
\begin{equation}
    \begin{split}
        \mathrm{M} &\in \left[2400,3900\right)K \\
        \mathrm{K} &\in \left[3900,5300\right)K \\
        \mathrm{G} &\in \left[5300,6000\right)K \\
        \mathrm{F} &\in \left[6000,7300\right)K \\
        \mathrm{A} &\in \left[7300,10000\right)K.
    \end{split}
\end{equation}
For G type stars, the SAG13 group found the sample geometric mean and variance from all of the submissions in each bin of the period-radius grid. They then performed three least-squares fits of joint power laws broken by radius of the form
\begin{equation}
    \frac{\partial^2 \eta}{\partial \ln R_p \partial \ln P} = C_i R_p^{\beta_i} P^{\alpha_i}.
\end{equation} 
The mean occurrence rate values were used to generate a ``baseline'' fit, the mean minus the standard deviation generated a ``pessimistic'' fit, and the mean plus the standard deviation generated an ``optimistic'' fit. The break between the two pieces of the power law was chosen to be 3.4$ R_\oplus $ to align with \cite{burke2015terrestrial} and for similar reasons. The parameter values found by least-squares are shown in Table~\ref{tab:SAG13} where the set of $ C_0 $, $ \alpha_0 $, and $ \beta_0 $ correspond to the parameters for $ R_p < 3.4R_\oplus $ and the other parameters correspond to $ R_p \geq 3.4R_\oplus $. It should be noted that the SAG13 power laws may not hold if extrapolated to regions where Kepler has poor reliability and completeness, including long period and/or small planets. In particular, a power law in period with $\alpha>0$ will violate the Hill stability criterion for a large enough period, so any extrapolation must either truncate the power law for some large $P$, or set $\alpha = 0$.

\begin{table}[ht]
    \caption[Parameter values for the three fits generated by the SAG13 group]{Parameter values for the three fits generated by the SAG13 group. $ \Omega_0 $, $ \alpha_0 $, and $ \rho_0 $ are for $ R_p < 3.4R_\oplus $ and $ \Omega_1 $, $ \alpha_1 $, and $ \rho_1 $ are for $ R_p \geq 3.4R_\oplus $.}
    \centering
    \begin{tabular}{cccc}
        \hline\hline
        & Pessimistic & Baseline & Optimistic \\
        \hline
        $ \Omega_0 $ & 0.138 & 0.38 & 1.06 \\
        $ \Omega_1 $ & 0.72 & 0.73 & 0.78 \\
        $ \alpha_0 $ & 0.204 & 0.26 & 0.32 \\
        $ \alpha_1 $ & 0.51 & 0.59 & 0.67 \\
        $ \rho_0 $ & 0.277 & -0.19 & -0.68 \\
        $ \rho_1 $ & -1.56 & -1.18 & -0.82 \\
        \hline
    \end{tabular}
    \label{tab:SAG13}
\end{table}

The likelihood function we use for our model fits requires knowledge of the occurrence rate uncertainty. Many of the data tables from SAG13 do not include uncertainty information on the occurrence rates, so we use the subset of SAG13 data tables that includes occurrence rate 1$ \sigma $ uncertainty information. We select the SAG13 data from the folders ``Natalie9p1'' (hereafter ``Batalha'', asymmetric uncertainties for M, K, and G type stars), ``Mulders'' (symmetric uncertainties for M, K, G, and F type stars), and ``Burke'' (asymmetric uncertainties for M and GK type stars). ``Batalha,'' contributed by Natalie Batalha, contains unpublished occurrence rate tables using the Q1-Q16 Kepler planet catalog \citep{mullally2015planetary}, DR24 star properties \citep{huber2014revised}, and the completeness calculation of \cite{christiansen2015measuring} with the analytic approximation to the window function given by \cite{burke2015terrestrial}. ``Mulders,'' contributed by Gijs Mulders, also uses the Q1-Q16 Kepler planet catalog \citep{mullally2015planetary}, DR24 star properties \citep{huber2014revised}, and the completeness calculation of \cite{christiansen2015measuring} as described in \cite{mulders2015increase} and \cite{mulders2015stellar}. ``Burke,'' contributed by Chris Burke, follows \cite{burke2015terrestrial} in using the Q1-Q16 Kepler planet catalog \citep{mullally2015planetary} and pipeline completeness model from \cite{christiansen2015measuring}. We do not include published data from \cite{fressin2013false} or \cite{christiansen2015measuring} because the $ T_{eff} $ range in those studies combine F, G, and K type stars. We use the data from \cite{dressing2015occurrence} for M type stars as independent test data and do not include it in the data set for model fitting.

We note that we selected community-sourced data from SAG13 using Kepler DR24 results and do not use the most recent and comprehensive Kepler planet catalog \citep{thompson2018planetary}, revised stellar properties \citep{mathur2017revised,berger2018revised}, Kepler pipeline \citep{jenkins2017kepler}, detection efficiencies \citep{christiansen2017planet}, or Robovetter vetting process \citep{coughlin2017planet} from DR25. \cite{kopparapu2018exoplanet} used the planet list from \cite{thompson2018planetary}, stellar properties from \cite{mathur2017revised}, and completeness calculations with \texttt{KeplerPORT} \citep{burke2017kepler} to compare DR25 based occurrence rates with SAG13 occurrence rates. The SAG13 occurrence rates (Table 3 from \cite{kopparapu2018exoplanet}) and DR25 based occurrence rates (Table 4 from \cite{kopparapu2018exoplanet}) are consistent and the DR25 occurrence rates are within the uncertainties of the baseline SAG13 occurrence rates. Because the data we selected comes from SAG13, the comparison done by \cite{kopparapu2018exoplanet} shows that our model fits will be within the uncertainty of Kepler DR25 based occurrence rates.

We assume that the uncertainties on occurrence rates are symmetric and Gaussian so that we may use a Gaussian likelihood function. When the uncertainties on occurrence rates from the data are asymmetric $ \left(\Delta_+ \neq \Delta_- \right) $, we use data where the uncertainties are within 20\% of each other $ \left(\left\vert \Delta_+ - \Delta_-\right\vert\right)/\left(\frac{1}{2}\left(\Delta_+ + \Delta_- \right)\right) < 0.2 $ \citep{chen2016probabilistic}. We take the average of the remaining asymmetric uncertainties \citep{weiss2014mass,wolfgang2016probabilistic} to give the standard deviation of the occurrence rates. We find no significant difference in our model fits when using data where the asymmetric uncertainties are within 10\% or 20\%, however, using the 20\% limit includes more data.  We also remove cases where the lower bound on the occurrence rate uncertainty would result in an occurrence rate of zero. These cuts remove 13\% of the data but result in symmetric uncertainties which may be used with a Gaussian likelihood function.

\begin{figure}[ht]
\figurenum{1}
\label{fig:Teffdata}
    \plotone{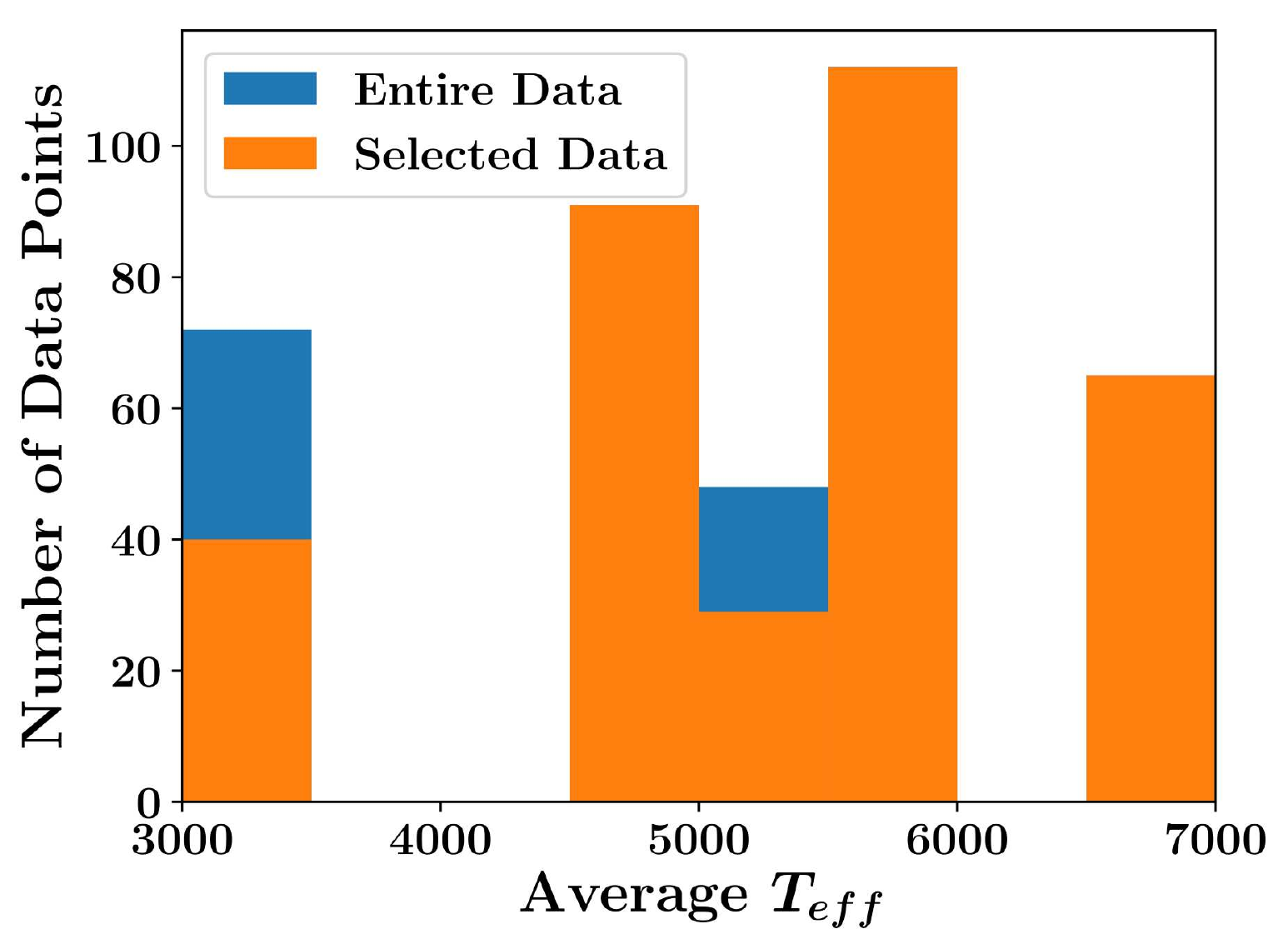}
    \caption{Distributions of the average $ T_{eff} $ for the entire and selected sets of data from SAG13 sources ``Batalha,'' ``Mulders,'' and ``Burke.''}
\end{figure}

\begin{figure}
\figurenum{2}
\label{fig:planetdata}
    \plotone{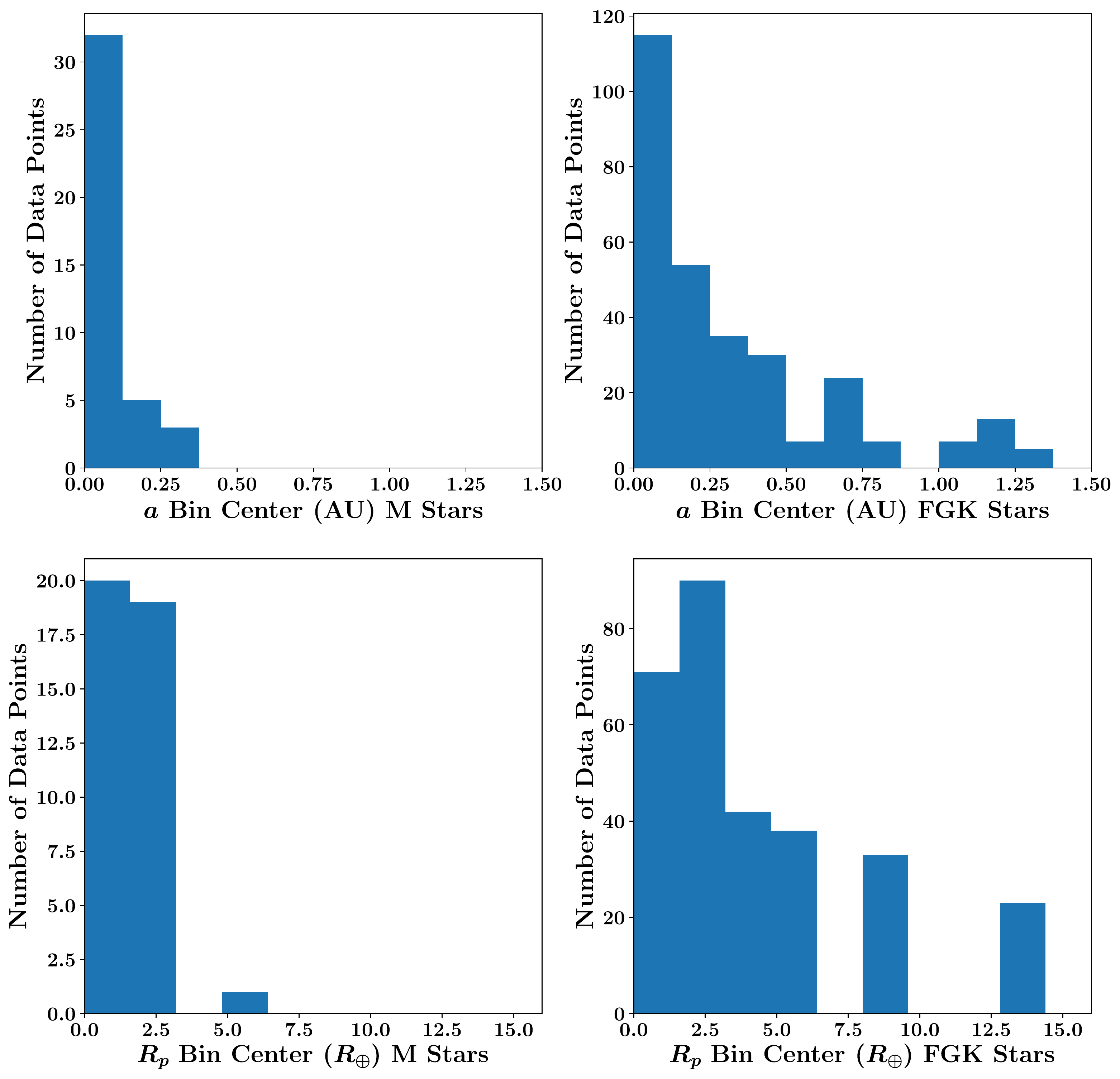}
    \caption{Distributions of the selected data from SAG13 sources ``Batalha,'' ``Mulders,'' and ``Burke.'' The top row shows the distribution of semi-major axis (M type stars on the left, F, G, and K type stars on the right). The bottom row shows the distribution of planet radius (M type stars on the left, F, G, and K type stars on the right).}
\end{figure}

For each of the occurrence rate data tables grouped by stellar spectral type in ``Batalha,'' ``Mulders,'' and ``Burke,'' we take the arithmetic average of the $ T_{eff} $ bin edges given in the meta-data, or take the arithmetic average of the SAG13 standard stellar spectral type bin edges when not given in the meta-data, and use these average $ T_{eff} $ values as inputs to our occurrence rate density models. Figure~\ref{fig:Teffdata} shows the distribution of average $ T_{eff} $ for the entire data set and the selected data set.  For each data set, we convert period bins to semi-major axis bins by finding the average stellar mass using the mass-$ T_{eff} $ relation from \cite{pecaut2013intrinsic}\footnote{\url{http://www.pas.rochester.edu/~emamajek/EEM_dwarf_UBVIJHK_colors_Teff.txt}}. Figure~\ref{fig:planetdata} shows the distributions of the selected semi-major axis and planet radius split between M type stars and F, G, and K type stars. Table~\ref{tab:dataset} provides an overall summary of the entire data set and selected data. Since Kepler's objective is to determine the frequency of Earth-sized planets in the habitable zone of sun-like stars \citep{batalha2014exploring,borucki2017kepler}, it is unsurprising that our selected data is concentrated near sun-like stars with small, close-in planets.

\begin{table}[ht]
    \caption{Overview of occurrence rate data from SAG13 sources. In the data points column, the first number is the total number of data points from the given source and the number in parenthesis is the number of data points used to fit our models.}
    \centering
    \begin{tabular}{cccccc}
        \hline\hline
        Source & $ T_{eff} $ (K) & Spectral Type & $ a $ (AU) & $ R_p \; \left(R_\oplus\right) $ & Data Points \\
        \hline
        \multirow{3}{*}{Batalha} & $ \left[2400,3900\right] $ & M & $ \left[0.059,0.941\right] $ & $ \left[0.67,17\right] $ & 10 (10) \\
        & $ \left[3900,5300\right] $ & K & $ \left[0.082,1.314\right] $ & $ \left[0.67,17\right] $ & 34 (34) \\
        & $ \left[5300,6000\right] $ & G & $ \left[0.091,1.449\right] $ & $ \left[0.67,17\right] $ & 40 (40) \\
        \hline
        \multirow{4}{*}{Mulders} & $ \left[2400,3900\right] $ & M & $ \left[0.009,0.941\right] $ & $ \left[0.44,26\right] $ & 27 (27) \\
        & $ \left[3900,5300\right] $ & K & $ \left[0.013,1.314\right] $ & $ \left[0.44,26\right] $ & 57 (57) \\
        & $ \left[5300,6000\right] $ & G & $ \left[0.014,1.449\right] $ & $ \left[0.44,26\right] $ & 72 (72) \\
        & $ \left[6000,7300\right] $ & F & $ \left[0.016,1.615\right] $ & $ \left[0.44,26\right] $ & 65 (65) \\
        \hline
        \multirow{2}{*}{Burke} & $ \left[2400,4200\right] $ & M & $ \left[0.063,0.634\right] $ & $ \left[0.67,11\right] $ & 35 (3) \\
        & $ \left[4200,6100\right] $ & GK & $ \left[0.087,1.387\right] $ & $ \left[0.67,17\right] $ & 48 (29) \\
        \hline
        Total & & & & & 388 (337) \\
        \hline
    \end{tabular}
    \label{tab:dataset}
\end{table}

\section{Fitting Planet Occurrence Rate Models}\label{sec:fits}
\begin{figure}[ht]
\figurenum{3}
\label{fig:HBM}
    \epsscale{0.5}
    \plotone{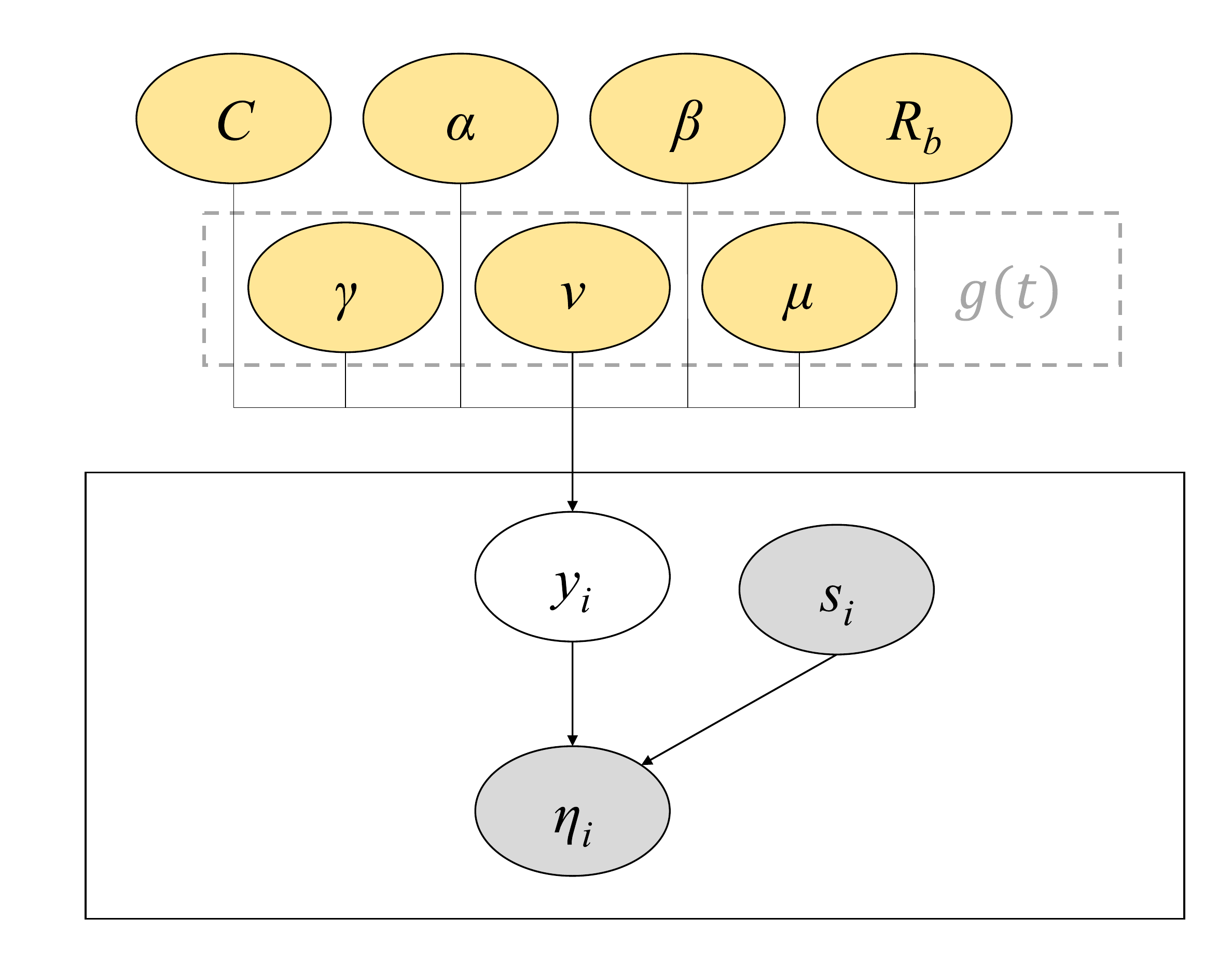}
    \caption{Graphical model of the hierarchical structure used to determine the occurrence rate density models in this work. Yellow ovals represent the model parameters (see Section~\ref{sec:models}), white ovals represent the true occurrence rates, and gray ovals represent data inputs. The parameters in the gray dashed box belong to the function $ g\left(t\right) $ and some or all of these values may be used in each model.}
\end{figure}

We use Bayesian parameter estimation with Markov Chain Monte Carlo (MCMC) sampling to evaluate the posterior distribution of our model parameters via the Python package \texttt{emcee} \citep{foreman2013emcee}\footnote{\url{https://github.com/dfm/emcee}}. The hierarchical model for our occurrence rate density model is shown in Figure~\ref{fig:HBM}. To sample from the posterior distribution, we require a likelihood function and prior. We assume a Gaussian likelihood function where the log-likelihood is given by
\begin{equation}
    \widehat{\mathcal{L}} = -\frac{1}{2}\sum_{i=1}^N \left[ \ln\left(2\pi s_i^2\right) + \frac{\left(\eta_i - y_i\right)^2}{s_i^2} \right]
\end{equation}
where $ \eta_i $ is the occurrence rate from the data, $ s_i $ is the 1$ \sigma $ error bar from the data, and
\begin{equation}
    y_i = \int_{\ln{R_{p,l_i}}}^{\ln{R_{p,u_i}}} \int_{\ln{a_{l_i}}}^{\ln{a_{u_i}}} \frac{\partial^2 \eta}{\partial \ln a \partial \ln R_p} d \ln{a} d \ln{R_p}
\end{equation}
is the occurrence rate determined from our model for occurrence rate density integrated over the lower $ \left(R_{p,l}, a_l\right) $ and upper $ \left(R_{p,u}, a_u\right) $ bin edges from the data. Our priors for both the simple and break radius models are given by
\begin{equation}
    \begin{split}
        \ln{C} &\sim \mathcal{U}\left(-5,10\right) \\
        \alpha &\sim \mathcal{U}\left(-2,2\right) \\
        \beta &\sim \mathcal{U}\left(-2,2\right) \\
        R_b &\sim \mathcal{U}\left(0.44,26\right) \\
        \gamma &\sim \mathcal{U}\left(-100,100\right) \\
        \mu &\sim \mathcal{U}\left(-500,500\right) \\
        \nu &\sim \mathcal{U}\left(-5000,5000\right)
    \end{split}
\end{equation}
where $ \mathcal{U} $ denotes a uniform distribution. Occurrence rates by definition are non-negative, so we sample $ \ln{C} $ to ensure this constraint. The power law indices for period and planetary radius from the literature tend to fall in the range of (-1,1), so we expand these limits to (-2,2) in our prior distributions. We set the limits of the $ R_b $ prior to the full range of $ R_p $ in our data set. We have no knowledge of the constants in $ g\left(t\right) $, so the limits on their prior distributions are wide. The values of $ t $ from our data are limited to (-0.6,0.3) and the $ \gamma $ (linear coefficient), $ \mu $ (quadratic coefficient), and $ \nu $ (cubic coefficient) terms have increasingly wide prior distribution limits.

We first perform a $ \chi^2 $ minimization for the model parameters to find initial values for the walkers in \texttt{emcee} \citep{foreman2013emcee}. For each set of posterior samples, we run 500 walkers for each parameter. The walkers must be initialized with differing values \citep{goodman2010ensemble,foreman2013emcee}, so we initialized each walker in the neighborhood of the $ \chi^2 $ minimization result by taking the $ \chi^2 $ minimization result and adding $ 10^{-4} $ times a random sample of a zero mean Gaussian with standard deviation of one. We discard a ``burn-in'' set of samples of 1000 steps and run 1000 additional steps to generate the final set of 500,000 samples for each parameter. The code generating this data, ``burn-in'' samples, and final posterior samples can be found in the github repository dgarrett622/Occurrence.

\section{Results}\label{sec:results}
To evaluate the goodness of fit for our models, we report values of $ \widehat{\mathcal{L}} $, $ \chi^2 $, and the Bayesian Information Criterion (BIC) \citep{schwarz1978estimating},
\begin{equation}\label{eq:BIC}
    \mathrm{BIC} = \ln{\left(N\right)}k - 2\widehat{\mathcal{L}},
\end{equation}
evaluated at the median value of each parameter, where $ N $ is the total number of data points and $ k $ is the number of parameters in the model. The model with the lowest BIC is preferred and differences larger than 10 are very strong, six to 10 are strong, two to six are positive, and zero to two are not significant \citep{kass1995bayes}. We note that BIC performs better as a selection metric when $ N >> k $ \citep{schwarz1978estimating}. When the number of model parameters is large, BIC has the potential to give a lower value due to over-fitting.

We first investigate the simple models with no break radius fit to all of the selected data. We find the model fit parameters for the simple model with no dependence on $ t $ (SMAll), linear dependence on $ t $ (SMtAll), quadratic dependence on $ t $ (SMt2All), and cubic dependence on $ t $ (SMt3All). We report the 16th, 50th, and 84th percentiles for each of the parameters from their marginalized distributions as well as $ \widehat{\mathcal{L}} $, $ \chi^2 $, and BIC goodness of fit values for the 50th percentile model parameter values in Table~\ref{tab:simpleall}. It is clear from the reduction in $ \chi^2 $ and BIC from SMAll to SMtAll that including $ g\left(t\right) $ in the model results in a better fit. From Table~\ref{tab:simpleall}, it appears that including more terms in $ g\left(t\right) $ also results in better fits to the data. 

\begin{table}[ht]
    \caption{16th, 50th, and 84th percentile model parameters from their marginalized distributions for simple models fit to all of the selected data. Goodness-of-fit values are evaluated at the 50th percentile model parameters.}
    \centering
    \begin{tabular}{ccccc}
        \hline\hline
        & SMAll & SM$\tau$All & SM$\tau$2All & SM$\tau$3All  \\
        \hline 
        $ \ln C $ & $ 0.467_{-0.039}^{+0.038} $ & $ 0.382_{-0.039}^{+0.038} $ & $ 0.338_{-0.043}^{+0.042} $ & $ 0.383_{-0.044}^{+0.043} $ \\
        $ \alpha $ & $ 1.245_{-0.012}^{+0.012} $ & $ 1.169_{-0.012}^{+0.012} $ & $ 1.171_{-0.012}^{+0.012} $ & $ 1.169_{-0.012}^{+0.012} $ \\
        $ \rho $ & $ -1.238_{-0.023}^{+0.023} $ & $ -1.215_{-0.022}^{+0.021} $ & $ -1.211_{-0.022}^{+0.021} $ & $ -1.195_{-0.022}^{+0.022} $ \\
        $ \lambda $ & & $ -2.961_{-0.129}^{+0.128} $ & $ -3.121_{-0.151}^{+0.148} $ & $ -1.965_{-0.305}^{+0.299} $ \\
        $ \omega $ & & & $ 2.951_{-1.219}^{+1.235} $ & $ -1.500_{-1.476}^{+1.526} $ \\
        $ \xi $ & & & & $ -32.93_{-7.742}^{+7.900} $ \\
        \hline
        $ \widehat{\mathcal{L}} $ & -407.6 & -150.1 & -147.2 & -138.9 \\
        $ \chi^2 $ & 3795 & 3280 & 3274 & 3257 \\
        $ \mathrm{BIC} $ & 832.7 & 323.5 & 323.4 & 312.9 \\
        \hline
    \end{tabular}
    \label{tab:simpleall}
\end{table}

\begin{figure}[ht]
\figurenum{4}
\label{fig:howardcompsimpleall}
    \plotone{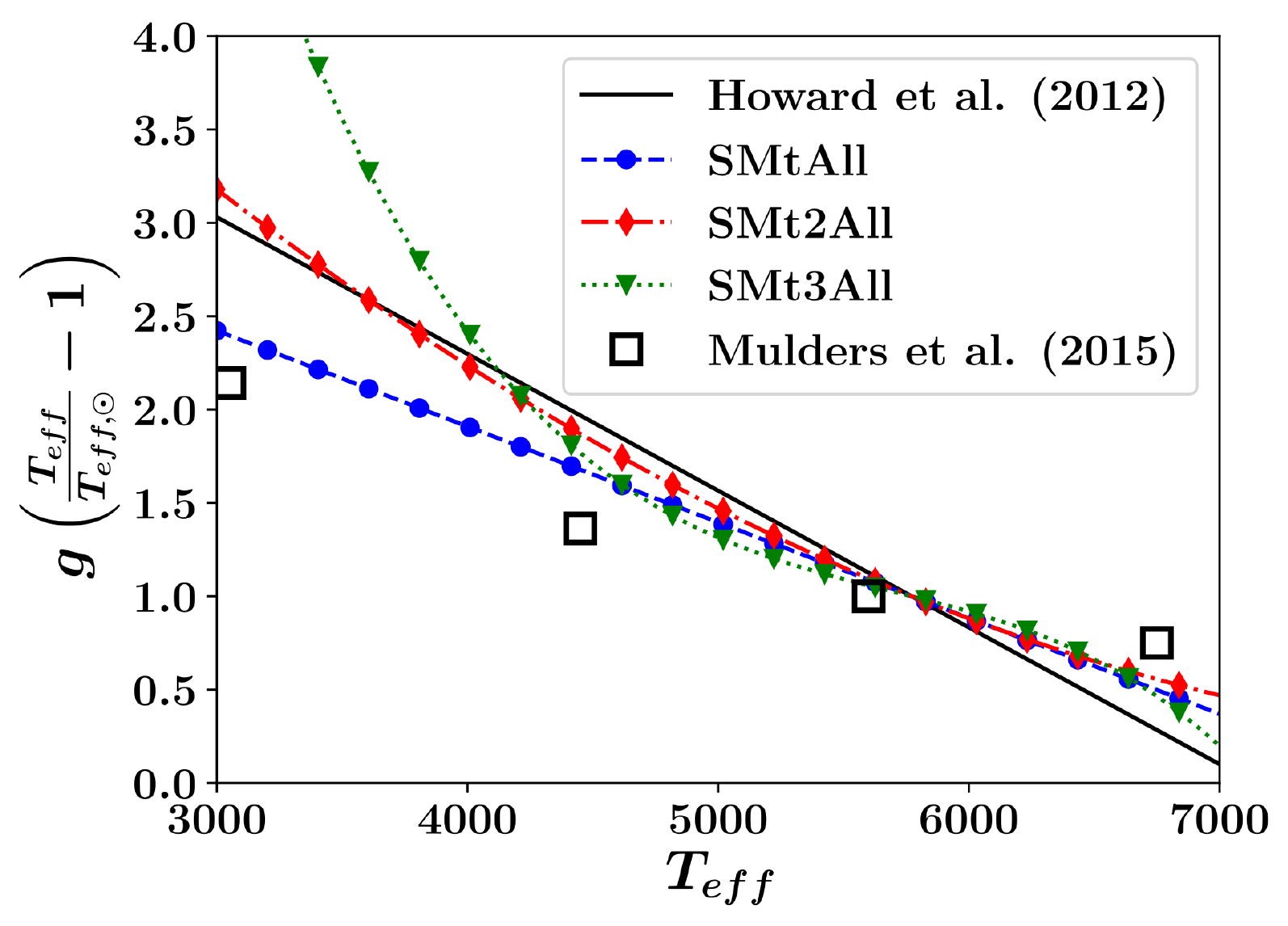}
    \caption{Simple models fit to all of the data compared to \cite{howard2012planet} and \cite{mulders2015stellar}. The models diverge for small $ T_{eff} $ (M type stars) but converge for $ T_{eff} > 4500 K $ (K, G, and F type stars). Because the SMt3All model has a higher potential for over-fitting and shows a large divergence from the SMtAll and SMt2All models and the data from \cite{howard2012planet} and \cite{mulders2015stellar}, it is not preferred. The SMtAll and SMt2All models have similar BIC values, but the SMtAll model is simpler and preferred.}
\end{figure}

To determine which of the simple models is the best fit to available data, we compare our fit results to \cite{howard2012planet} and \cite{mulders2015stellar} in Figure~\ref{fig:howardcompsimpleall}. We normalize Equation 9 from \cite{howard2012planet} (relating overall planet occurrence rates and $ T_{eff} $) such that it is 1 when $ t = 0 $. We take the values scaling the overall occurrence rates from Figure 4 of \cite{mulders2015stellar} (for 1--4$ R_\oplus $) and renormalize to G type stars to better represent our results normalized to solar $ T_{eff} $. \cite{howard2012planet} fit their Equation 9 to planet occurrence rate data for planets up to 0.25 AU and 2--4$R_{\oplus}$. The data for our model fits extend beyond their semi-major axis range and include a wider range for planetary radius, so it is unsurprising that our results give a different relation. However, as in Figure 8 of \cite{howard2012planet} we find that overall planet occurrence rates decrease with increasing $ T_{eff} $. SMtAll, SMt2All, and SMt3All diverge from each other for $ T_{eff} < 4500 K $ but converge for $ T_{eff} > 4500 K $. This indicates that a break in $ g\left(t\right) $ may facilitate a better fit to the data. 

\begin{figure}[ht]
\figurenum{5}
\label{fig:cornersimpleall}
    \plotone{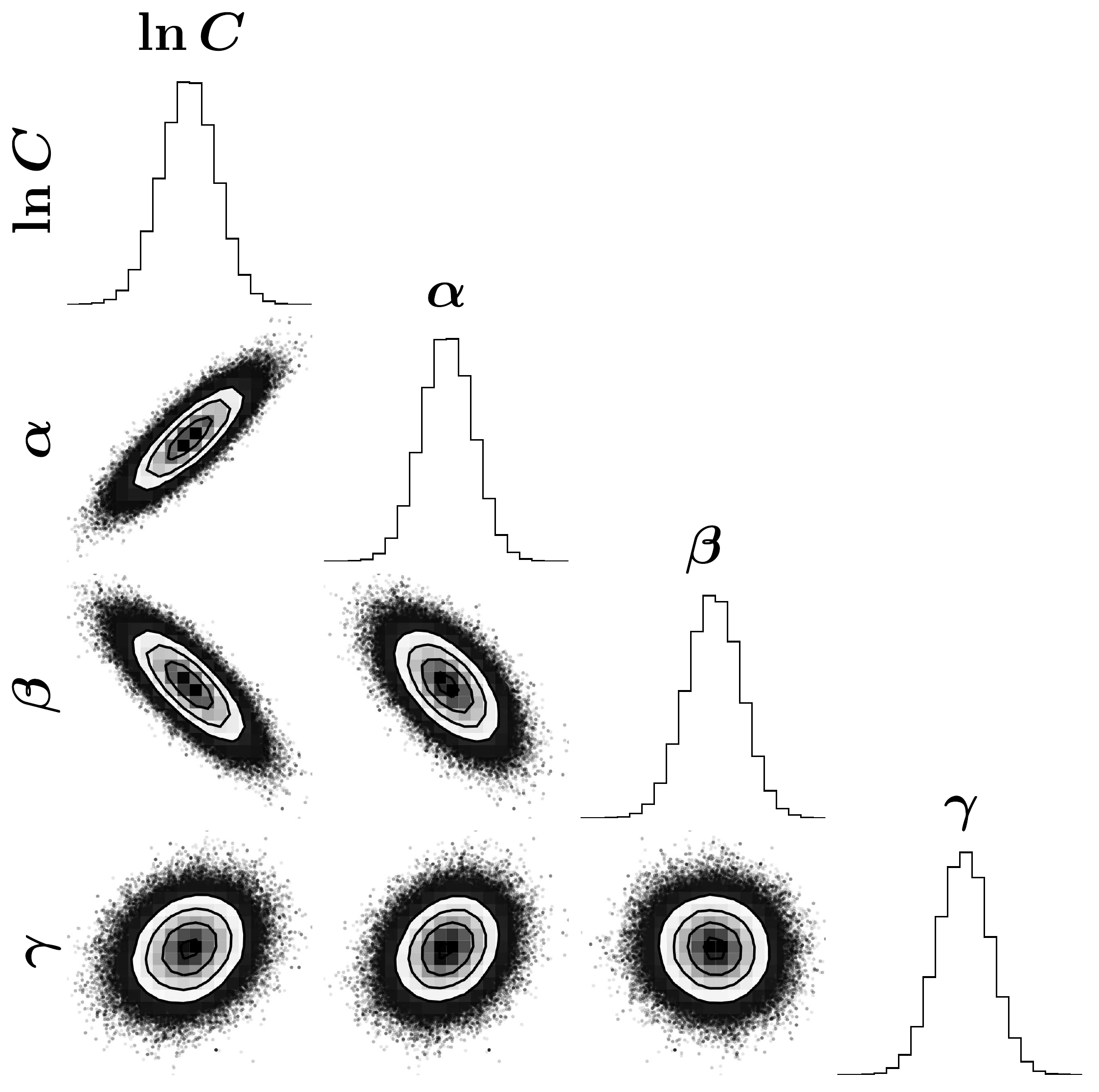}
    \caption{Corner plot for SMtAll model using \texttt{corner.py} \citep{foreman2016corner}.}
\end{figure}

Although it has the best BIC value (Table~\ref{tab:simpleall}), we do not consider the SMt3All model to be the preferred model because of the higher potential for over-fitting and large divergence from SMtAll and SMt2All (Figure~\ref{fig:howardcompsimpleall}). SMtAll and SMt2All are in the neighborhood of both \cite{howard2012planet} and \cite{mulders2015stellar} and have very nearly the same BIC. Because of this, we select the simpler SMtAll model as the preferred simple model fit to all of the selected data. We present a corner plot \citep{foreman2016corner}\footnote{\url{https://github.com/dfm/corner.py}} of the SMtAll model in Figure~\ref{fig:cornersimpleall}.

\begin{figure}
\figurenum{6}[ht]
\label{fig:howardcompbreakall}
    \epsscale{0.5}
    \plotone{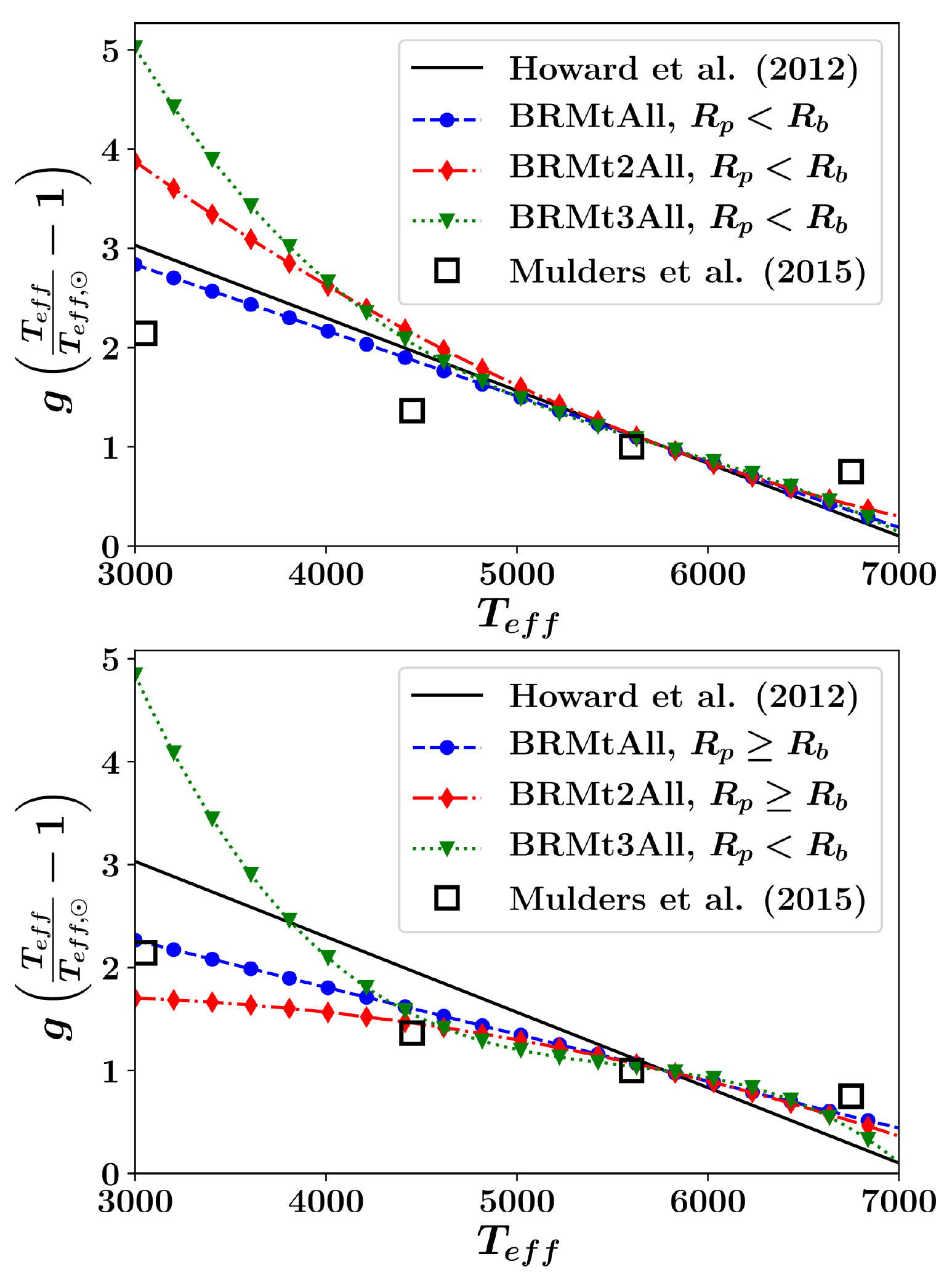}
    \caption{Comparison of break radius models fit to all of the selected data to \cite{howard2012planet} and \cite{mulders2015stellar}. The top panel shows the break radius models where $ R_p < R_b $, and the bottom panel shows the break radius models where $ R_p \geq R_b $. As with the simple models, the break radius models diverge for $ T_{eff} < 4500 K $ but converge for $ T_{eff} > 4500 K $. The BRMtAll model is preferred by BIC.}
\end{figure}

\begin{figure}[ht]
\figurenum{7}
\label{fig:cornerbreakall}
    \plotone{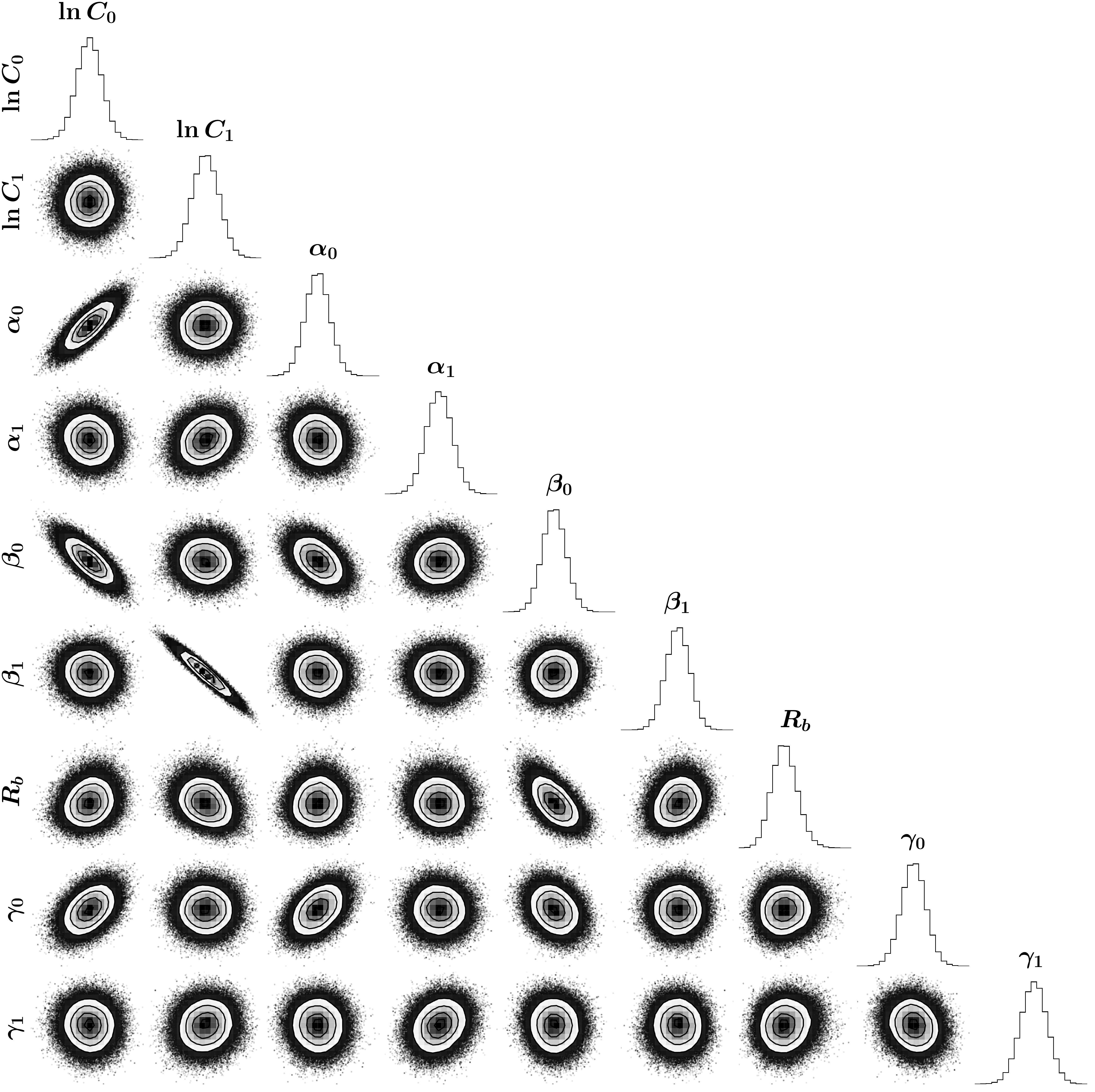}
    \caption{Corner plot for BRMtAll model using \texttt{corner.py} \citep{foreman2016corner}.}
\end{figure}

The break radius models fit to all of the selected data with no dependence on $ t $ (BRMAll), linear dependence on $ t $ (BRMtAll), quadratic dependence on $ t $ (BRMt2All), and cubic dependence on $ t $ (BRMt3All) follow similar trends to the simple models. We report the model parameter values at the 16th, 50th, and 84th percentile as well as the goodness of fit values evaluated at the model parameter 50th percentile values in Table~\ref{tab:breakall}. By any of the goodness of fit measures, the break radius model is a  better fit than the equivalent version of the simple model. As with the simple models, the break radius models fit the data better when $ g_i\left(t\right) $ is included. We compare the break radius model fits to \cite{howard2012planet} and \cite{mulders2015stellar} in Figure~\ref{fig:howardcompbreakall}. The top panel of Figure~\ref{fig:howardcompbreakall} shows our break radius models for $ R_p < R_b $ and we see the same trend of decreasing occurrence rates with increasing $ T_{eff} $ as \cite{howard2012planet}. We also note that the BRMtAll model closely reproduces the result from \cite{howard2012planet} for planets with $ R_p < 2.763R_\oplus $. \cite{howard2012planet} found that occurrence rates for planets larger than 4$ R_\oplus $ have no correlation with $ T_{eff} $. In the bottom panel of Figure~\ref{fig:howardcompbreakall} we find that occurrence rate decreases for increasing $ T_{eff} $ for planets larger than our break radius which does include some planets with $ R_p < 4 R_\oplus $, so we cannot make a direct comparison to the \cite{howard2012planet} finding. As with the simple models, we also see the divergence of BRMtAll, BRMt2All, and BRMt3All near 4500 K. We find that the break radius for the BRMAll model is 3.027$ R_\oplus $ which compares favorably to SAG13 (3.4$ R_\oplus $) and \cite{burke2015terrestrial} (3.3$ R_\oplus $) results. From Figure~\ref{fig:howardcompbreakall} and Table~\ref{tab:breakall}, the model which best matches the \cite{mulders2015stellar} result and is preferred by BIC is BRMtAll. We present the corner plot of the BRMtAll model in Figure~\ref{fig:cornerbreakall}

\begin{table}
    \caption[Model parameters for break radius models fit to all of the selected data]{16th, 50th, and 84th percentile model parameters from their marginalized distributions for break radius models fit to all of the selected data. Goodness-of-fit values are evaluated at the 50th percentile model parameters.}
    \centering
    \begin{tabular}{ccccc}
        \hline\hline
        & BRMAll & BRM$\tau$All & BRM$\tau$2All & BRM$\tau$3All \\
        \hline
        $ \ln \Omega_0 $ & $ 0.412_{-0.045}^{+0.042} $ & $ 0.020_{-0.053}^{+0.051} $ & $ -0.047_{-0.057}^{+0.056} $ & $ 0.00483_{-0.060}^{+0.057} $ \\
        $ \ln \Omega_1 $ & $ -0.453_{-0.211}^{+0.197} $ & $ -0.638_{-0.187}^{+0.186} $ & $ -0.566_{-0.204}^{+0.203} $ & $ -0.541_{-0.204}^{+0.202} $ \\
        $ \alpha_0 $ & $ 1.266_{-0.012}^{+0.012} $ & $ 1.096_{-0.013}^{+0.013} $ & $ 1.096_{-0.013}^{+0.013} $ & $ 1.100_{-0.013}^{+0.013} $ \\
        $ \alpha_1 $ & $ 1.035_{-0.044}^{+0.036} $ & $ 1.004_{-0.028}^{+0.029} $ & $ 1.005_{-0.028}^{+0.028} $ & $ 1.005_{-0.027}^{+0.028} $ \\
        $ \rho_0 $ & $ -0.676_{-0.107}^{+0.096} $ & $ -0.169_{-0.070}^{+0.073} $ & $ -0.142_{-0.070}^{+0.070} $ & $ -0.157_{-0.070}^{+0.073} $ \\
        $ \rho_1 $ & $ -0.961_{-0.108}^{+0.106} $ & $ -0.880_{-0.101}^{+0.098} $ & $ -0.903_{-0.104}^{+0.103} $ & $ -0.892_{-0.101}^{+0.101} $ \\
        $ R_b $ & $ 3.027_{-0.142}^{+0.323} $ & $ 2.763_{-0.045}^{+0.049} $ & $ 2.738_{-0.046}^{+0.048} $ & $ 2.748_{-0.047}^{+0.051} $ \\
        $ \lambda_0 $ & & $ -3.826_{-0.145}^{+0.146} $ & $ -4.116_{-0.189}^{+0.0.186} $ & $ -3.308_{-0.385}^{+0.370} $ \\
        $ \omega_0 $ & & & $ 3.885_{-1.339}^{+1.412} $ & $ 0.828_{-1.732}^{+1.825} $ \\
        $ \xi_0 $ & & & & $ -19.99_{-8.487}^{+8.726} $ \\
        $ \lambda_1  $ & & $ -2.636_{-0.290}^{-0.290} $ & $ -2.526_{-0.349}^{+0.324} $ & $ -1.530_{-1.023}^{+0.934} $ \\
        $ \omega_1 $ & & & $ -2.190_{-2.759}^{+3.021} $ & $ -4.488_{-3.429}^{+3.833} $ \\
        $ \xi_1 $ & & & & $ -36.07_{-31.57}^{+34.35} $ \\
        \hline
        $ \widehat{\mathcal{L}} $ & -253.9 & 111.6 & 116.2 & 120.0 \\
        $ \chi^2 $ & 3487 & 2756 & 2747 & 2739 \\
        $ \mathrm{BIC} $ & 548.6 & -170.9 & -168.3 & -164.4 \\
    \end{tabular}
    \label{tab:breakall}
\end{table}

\begin{figure}[ht]
\figurenum{8}
\label{fig:cornersimpleM}
    \plotone{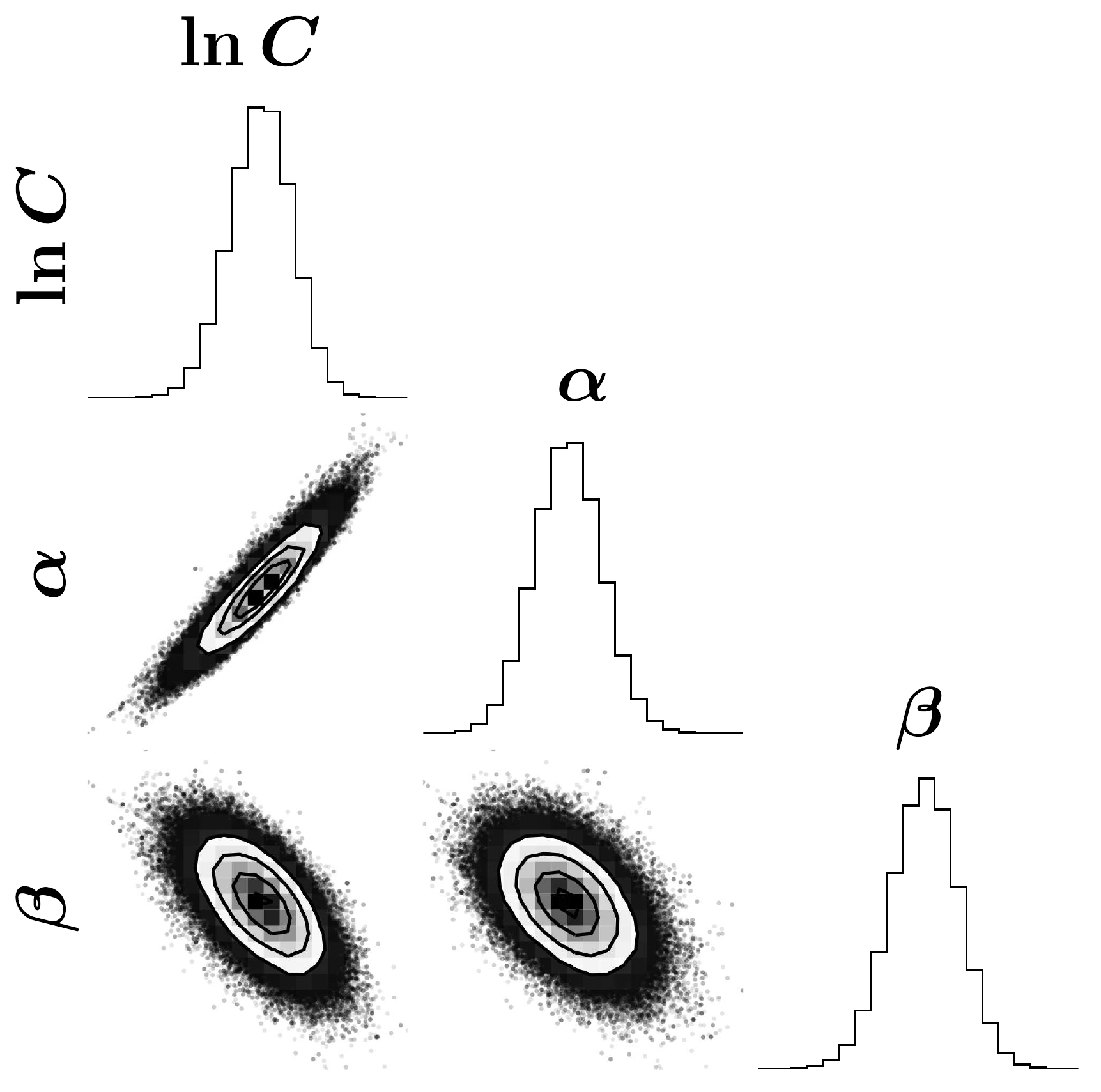}
    \caption{Corner plot for SMM model using \texttt{corner.py} \citep{foreman2016corner}.}
\end{figure}

We split the data to investigate the divergence at lower $ T_{eff} $ seen in Figure~\ref{fig:howardcompsimpleall} and Figure~\ref{fig:howardcompbreakall}. We split the data for the M type stars from the data for the F, G, and K type stars. For the M type star sample, we are left with SAG13 data from ``Batalha'' (10 data points), ``Mulders'' (27 data points), and ``Burke'' (3 data points). Unfortunately, using the simple arithmetic mean of the $ T_{eff} $ bin edges from the meta-data results in the ``Batalha'' and ``Mulders'' data sets having the same average $ T_{eff} $ while the ``Burke'' data set has a different average $ T_{eff} $ but only three data points (as seen in Table~\ref{tab:dataset}). Because of this, we do not have sufficient $ T_{eff} $ data points to perform fits including $ g\left(t\right) $ for the M type star data set alone. We perform a simple model fit (SMM) to get the model parameter values at the 16th, 50th, and 84th percentile of their marginalized distributions and goodness of fit values as $ \ln C = 2.483_{-0.240}^{+0.226} $, $ \alpha = 1.260_{-0.072}^{+0.073} $, $ \beta = -0.623_{-0.154}^{+0.154} $, $ \widehat{\mathcal{L}} = 64.72 $, $ \chi^2 = 80.82 $, and BIC = -118.4. We fit a break radius model to this data, however, there is no indication that a break radius model is better than the simple model. We note that the data for M type stars included in our model fitting contain very few planets larger than the break radii ($ \sim2.7-3R_\oplus $) determined by the model fits using the M, K, G, and F type star data (see Figure~\ref{fig:planetdata}). This is likely the reason that a break radius model is not preferred over the simple model. A corner plot for the SMM model is given in Figure~\ref{fig:cornersimpleM}.

\begin{figure}[ht]
\figurenum{9}
\label{fig:howardcompsimpleFGK}
    \plotone{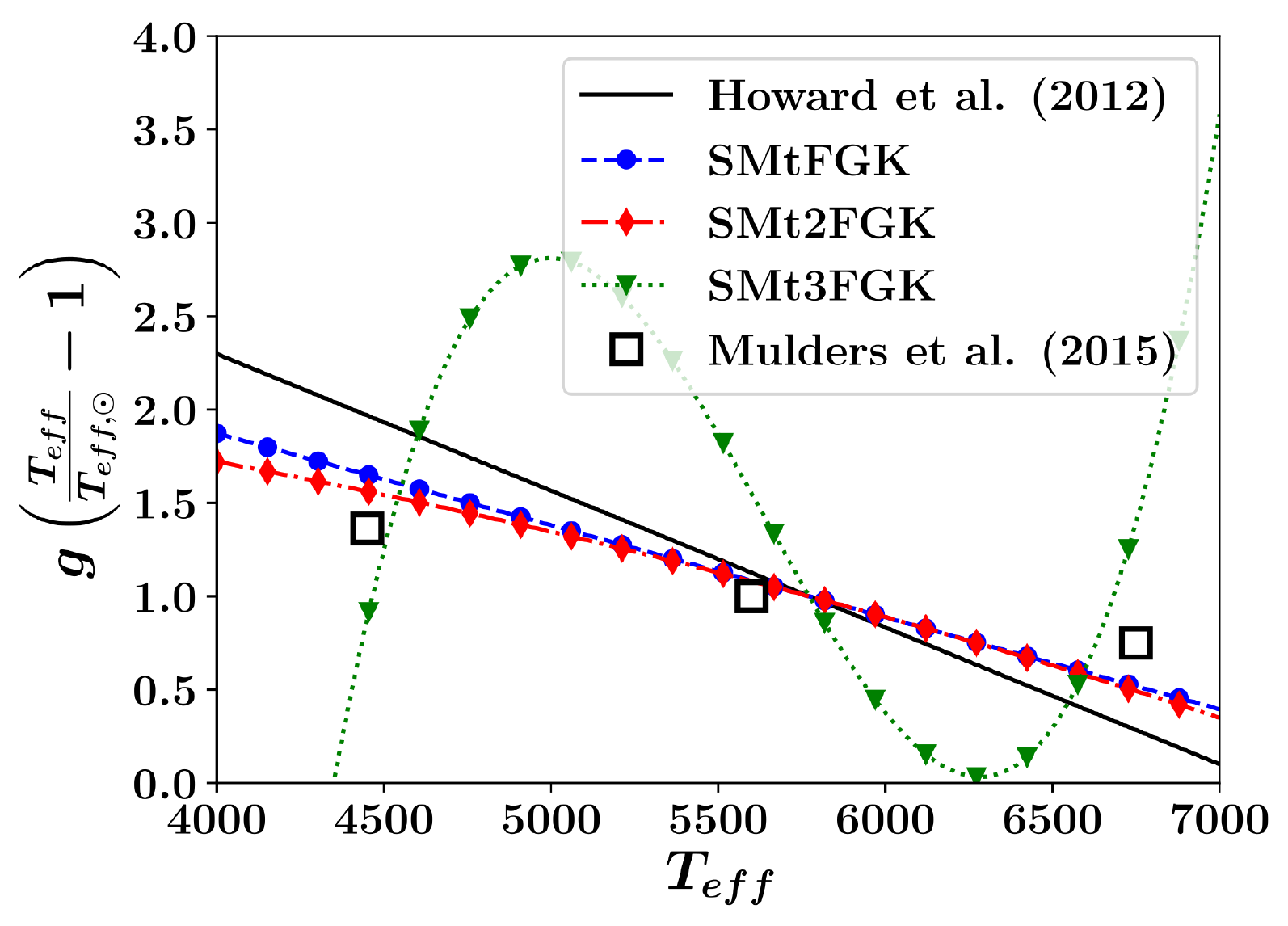}
    \caption{Comparison of simple models fit to the F, G, and K type star data to \cite{howard2012planet} and \cite{mulders2015stellar}. The SMtFGK and SMt2FGK models are in close agreement. Because the SMt3FGK model has a higher potential for over-fitting and shows large divergence from the other models and the data from \cite{howard2012planet} and \cite{mulders2015stellar}, it is not preferred. The SMtFGK model is then preferred by BIC.}
\end{figure}

\begin{figure}
\figurenum{10}[ht]
\label{fig:cornersimpleFGK}
    \plotone{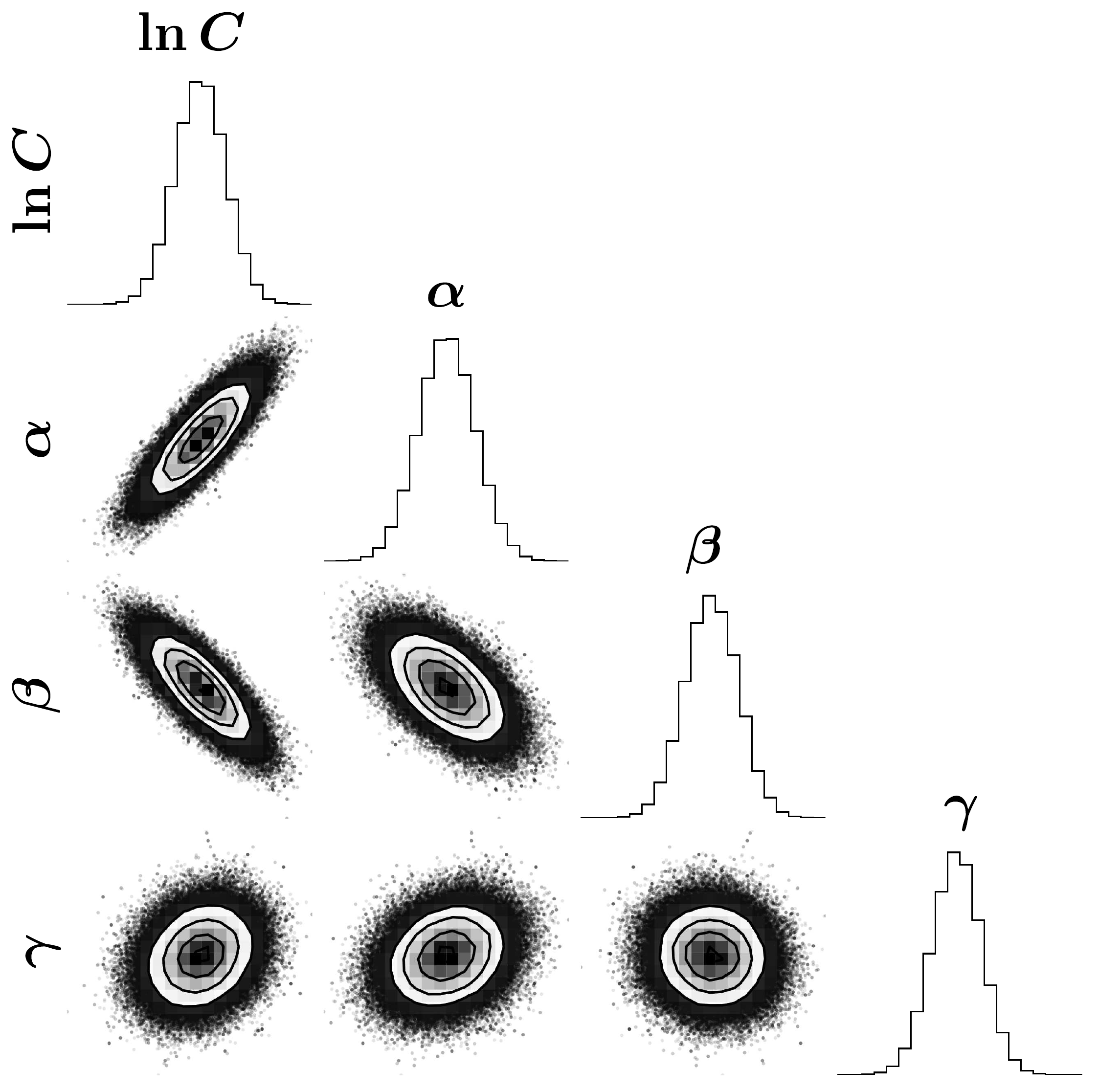}
    \caption{Corner plot for SMtFGK model using \texttt{corner.py} \citep{foreman2016corner}.}
\end{figure}

We present results for the simple models fit to the F, G, and K type star data with no break radius and no dependence on $ t $ (SMFGK), linear dependence on $ t $ (SMtFGK), quadratic dependence on $ t $ (SMt2FGK), and cubic dependence on $ t $ (SMt3FGK). We report the 16th, 50th, and 84th percentiles for each of the model parameters from their marginalized distributions as well as $ \widehat{\mathcal{L}} $, $ \chi^2 $, and BIC in Table~\ref{tab:simpleFGK}. Similar trends are seen in the simple models for the F, G, and K type fits as for the simple models fit to the entire selected data set. We again plot the comparison to \cite{howard2012planet} and \cite{mulders2015stellar} in Figure~\ref{fig:howardcompsimpleFGK} to determine which model is most accurate. The SMtFGK and SMt2FGK models show close agreement. Just as with the simple models fit to all of the selected data, the SMt3FGK model has a higher potential for over-fitting and shows large divergence from the other models and the data from \cite{howard2012planet} and \cite{mulders2015stellar} (Figure~\ref{fig:howardcompsimpleFGK}). Because of this, we do not select the SMt3FGK model as the preferred model even though it has the best BIC value (Table~\ref{tab:simpleFGK}). Of the remaining models the one preferred by BIC is the SMtFGK model. We give a corner plot \citep{foreman2016corner} of the SMtFGK model in Figure~\ref{fig:cornersimpleFGK}.

\begin{table}[ht]
    \caption{16th, 50th, and 84th percentile model parameters from their marginalized distributions for simple models fit to the F, G, and K type star data. Goodness-of-fit values are evaluated at the 50th percentile model parameters.}
    \centering
    \begin{tabular}{ccccc}
        \hline\hline
        & SMFGK & SM$\tau$FGK & SM$\tau$2FGK & SM$\tau$3FGK \\
        \hline
        $ \ln \Omega $ & $ 0.462_{-0.040}^{+0.038} $ & $ 0.380_{-0.039}^{+0.038} $ & $ 0.400_{-0.043}^{+0.042} $ & $ 0.136_{-0.062}^{+0.059} $ \\
        $ \alpha $ & $ 1.246_{-0.012}^{+0.012} $ & $ 1.174_{-0.012}^{+0.012} $ & $ 1.173_{-0.012}^{+0.012} $ & $ 1.193_{-0.013}^{+0.013} $ \\ 
        $ \rho $ & $ -1.235_{-0.024}^{+0.023} $ & $ -1.212_{-0.022}^{+0.022} $ & $ -1.214_{-0.021}^{+0.022} $ & $ -1.380_{-0.026}^{+0.026} $ \\
        $ \lambda $ & & $ -2.841_{-0.129}^{+0.130} $ & $ -2.764_{-0.147}^{+0.142} $ & $ -17.90_{-2.290}^{+2.030} $ \\
        $ \omega $ & & & $ -1.377_{-1.166}^{+1.219} $ & $ 34.44_{-5.301}^{+5.993} $ \\
        $ \xi $ & & & & $ 501.5_{-65.85}^{+74.33} $ \\
        \hline
        $ \widehat{\mathcal{L}} $ & -395.2 & -164.6 & -163.9 & -96.50 \\
        $ \chi^2 $ & 3560 & 3098 & 3097 & 2962 \\
        $ \mathrm{BIC} $ & 807.5 & 352.0 & 356.3 & 227.2 \\
        \hline
    \end{tabular}
    \label{tab:simpleFGK}
\end{table}

\begin{figure}[ht]
\figurenum{11}
\label{fig:howardcompbreakFGK}
    \epsscale{0.5}
    \plotone{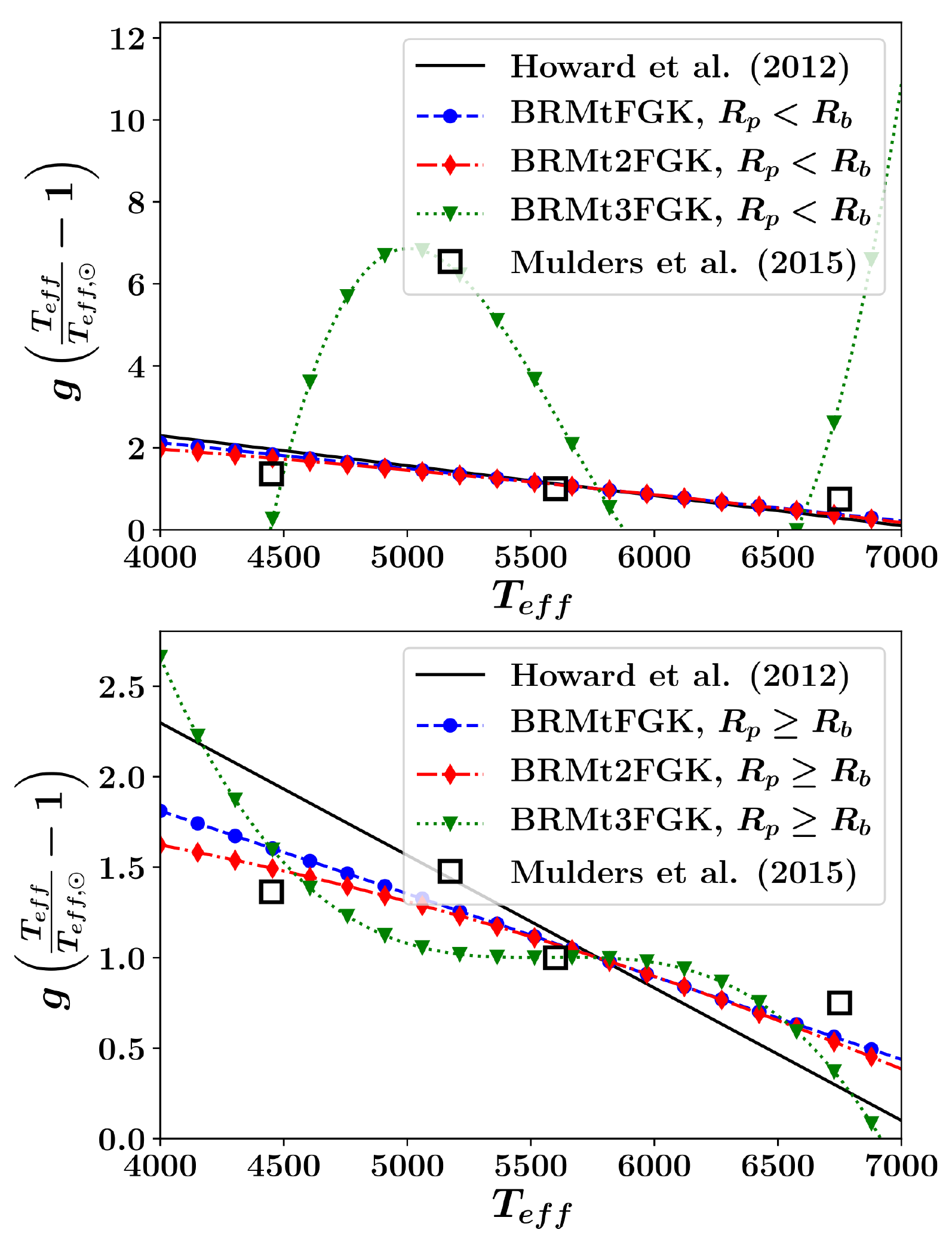}
    \caption{Comparison of break radius models to \cite{howard2012planet} and \cite{mulders2015stellar} using F, G, and K type star data. The top panel shows the break radius models where $ R_p < R_b $ and the bottom panel shows the break radius models where $ R_p \geq R_b $. The BRMtFGK and BRMt2FGK models reproduce the \cite{mulders2015stellar} result. Because of the higher potential for over-fitting and large divergence of the BRMt3FGK model from the other models and the data from \cite{howard2012planet} and \cite{mulders2015stellar}, it is not preferred. The BRMt2FGK model is very strongly preferred to the BRMtFGK model by BIC.}
\end{figure}

The break radius models fit to the F, G, and K type star data with no dependence on $ t $ (BRMFGK), linear dependence on $ t $ (BRMtFGK), quadratic dependence on $ t $ (BRMt2FGK), and cubic dependence on $ t $ (BRMt3FGK) follow similar trends to previous models. We report the model parameter values at the 16th, 50th, and 84th percentile values in Table~\ref{tab:breakFGK}. The break radius models are again better than the simple models fit to this data. We compare the break radius models for this data set to \cite{howard2012planet} and \cite{mulders2015stellar} in Figure~\ref{fig:howardcompbreakFGK}. The top panel of Figure~\ref{fig:howardcompbreakFGK} shows our break radius models for $ R_p < R_b $ and the bottom panel shows our break radius models for $ R_p \geq R_b $. The BRMtFGK and BRMt2FGK models closely reproduce the result from \cite{howard2012planet} for $ R_p < R_b $. We find the break radius for the BRMFGK model is 3.060$ R_\oplus $ which again compares favorably to SAG13 and \cite{burke2015terrestrial}. 

\begin{figure}[ht]
\figurenum{12}
\label{fig:cornerbreakFGK}
    \plotone{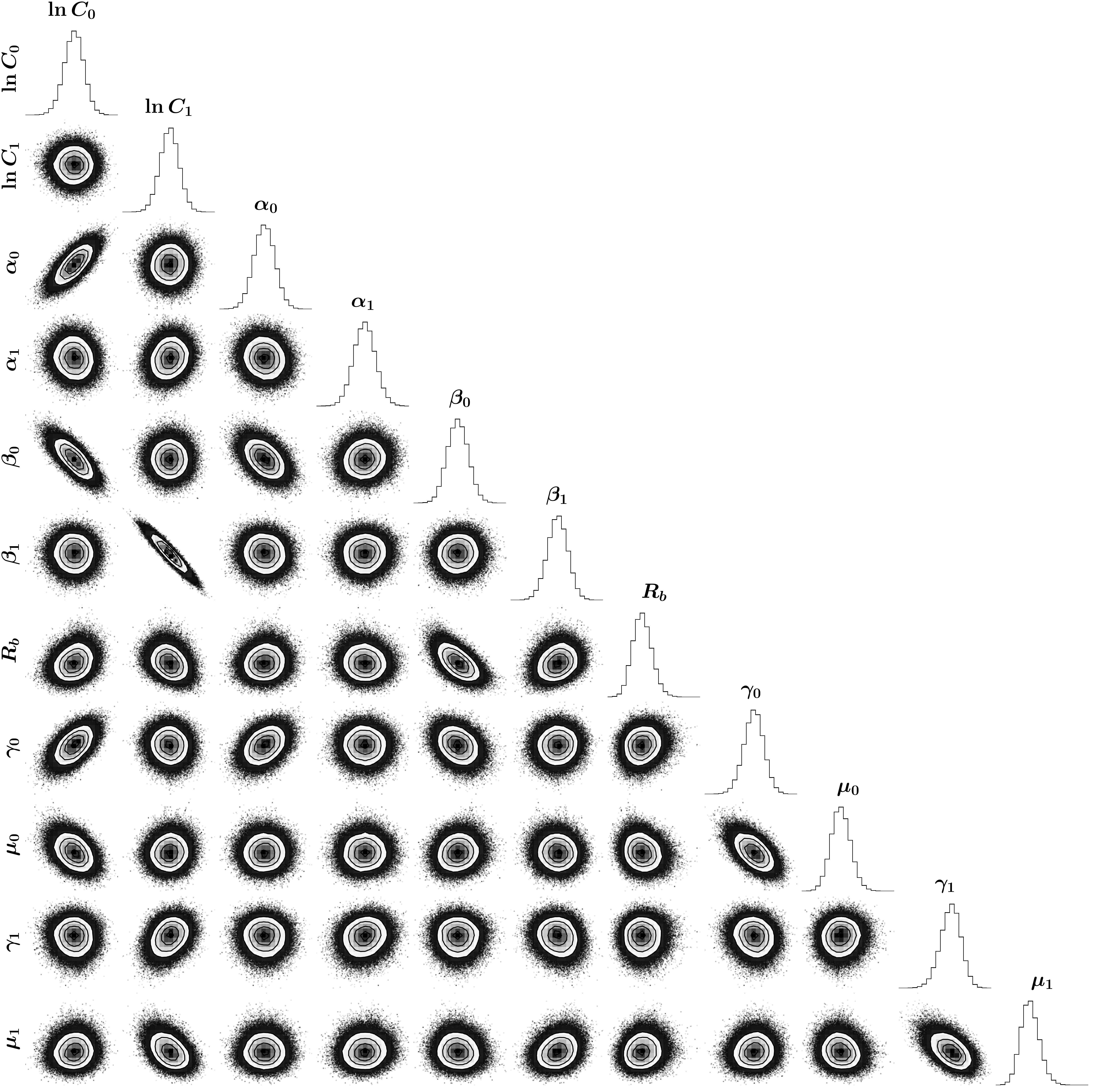}
    \caption{Corner plot for BRMt2FGK model using \texttt{corner.py} \citep{foreman2016corner}.}
\end{figure}

Even though the BIC value for the BRMt3FGK model is lower than the other models, we do not select it as the preferred model. As before, the BRMt3FGK model has a higher potential for over-fitting and exhibits large divergence from the other models and the data from \cite{howard2012planet} and \cite{mulders2015stellar}. Visually, the BRMtFGK and BRMt2FGK give comparable results, however, the BRMt2FGK model is very strongly preferred by BIC. We give a corner plot \citep{foreman2016corner} of the BRMt2FGK model in Figure~\ref{fig:cornerbreakFGK}.

\begin{table}
    \caption{Break radius model parameter 16th, 50th, and 84th percentile values fit to F, G, and K type star data. Goodness-of-fit values are evaluated at the 50th percentile model parameters.}
    \centering
    \begin{tabular}{ccccc}
        \hline\hline
        & BRMFGK & BRM$\tau$FGK & BRM$\tau$2FGK & BRM$\tau$3FGK \\
        \hline
        $ \ln \Omega_0 $ & $ 0.407_{-0.045}^{+0.043} $ & $ 0.027_{-0.052}^{+0.051} $ & $ 0.049_{-0.056}^{+0.056} $ & $ -0.936_{-0.148}^{+0.129} $ \\
        $ \ln \Omega_1 $ & $ -0.445_{-0.217}^{+0.200} $ & $ -0.629_{-0.190}^{+0.188} $ & $ -0.592_{-0.206}^{+0.203} $ & $ -0.214_{-0.202}^{+0.194} $ \\
        $ \alpha_0 $ & $ 1.268_{-0.012}^{+0.012} $ & $ 1.104_{-0.013}^{+0.013} $ & $ 1.104_{-0.013}^{+0.013} $ & $ 1.063_{-0.016}^{+0.016} $ \\
        $ \alpha_1 $ & $ 1.030_{-0.044}^{+0.037} $ & $ 1.006_{-0.028}^{+0.029} $ & $ 1.003_{-0.028}^{+0.029} $ & $ 1.007_{-0.025}^{+0.025} $ \\
        $ \rho_0 $ & $ -0.688_{-0.100}^{+0.104} $ & $ -0.175_{-0.071}^{+0.072} $ & $ -0.183_{-0.073}^{+0.074} $ & $ -0.263_{-0.068}^{+0.067} $ \\
        $ \rho_1 $ & $ -0.968_{-0.111}^{+0.110} $ & $ -0.884_{-0.101}^{+0.100} $ & $ -0.894_{-0.104}^{+0.103} $ & $ -1.022_{-0.103}^{+0.102} $ \\
        $ R_b $ & $ 3.060_{-0.157}^{+0.353} $ & $ 2.766_{-0.048}^{+0.052} $ & $ 2.771_{-0.049}^{+0.053} $ & $ 2.604_{-0.035}^{+0.037} $ \\
        $ \lambda_0 $ & & $ -3.676_{-0.148}^{+0.150} $ & $ -3.572_{-0.174}^{+0.171} $ & $ -57.95_{-10.66}^{+8.266} $ \\
        $ \omega_0 $ & & & $ -1.327_{-1.252}^{+1.315} $ & $ 124.6_{-20.04}^{+25.88} $ \\
        $ \xi_0 $ & & & & $ 1720_{-257.8}^{+332.3} $ \\
        $ \lambda_1 $ & & $ -2.642_{-0.296}^{+0.291} $ & $ -2.538_{-0.347}^{+0.325} $ & $ -0.219_{-0.962}^{+0.881} $ \\
        $ \omega_1 $ & & & $ -1.660_{-2.879}^{+3.177} $ & $ -8.129_{-2.895}^{+3.228} $ \\
        $ \xi_1 $ & & & & $ -81.54_{-30.86}^{+33.07} $ \\
        \hline
        $ \widehat{\mathcal{L}} $ & -242.6 & 84.74 & 85.58 & 263.7 \\
        $ \chi^2 $ & 3254 & 2600 & 2598 & 2242 \\
        $ \mathrm{BIC} $ & 525.0 & -118.2 & -108.5 & -453.3 \\ 
        \hline
    \end{tabular}
    \label{tab:breakFGK}
\end{table}

Visually comparing Figure~\ref{fig:howardcompbreakall} and Figure~\ref{fig:howardcompbreakFGK} show that the F, G, K type star data better reproduce the \cite{howard2012planet} and \cite{mulders2015stellar} results. Comparing Table~\ref{tab:breakall} and Table~\ref{tab:breakFGK} values show that the greatest differences in all of the parameters are in the $ g_i \left(t\right) $ coefficients. Comparing the SMM fit to the SMFGK fit and the SMAll fit shows a significant difference in the parameters. This indicates that the occurrence rate density function for M type stars must be different than F, G, and K type stars when considering only semi-major axis, planetary radius, and $ T_{eff} $. 

Our best-fit model is given as
\begin{equation}
    \frac{\partial^2 \eta}{\partial \ln a \partial \ln R_p} = 
    \begin{cases}
        11.98\left(\frac{a}{a_\oplus}\right)^{1.260}\left(\frac{R_p}{R_\oplus}\right)^{-0.623} & 2400 K \leq T_{eff} \leq 3900 K \\
        1.050\left[\left(\frac{a}{a_\oplus}\right)^{1.104}\left(\frac{R_p}{R_\oplus}\right)^{-0.183}\right]g_0\left(t\right) & R_p < 2.771R_\oplus, \quad 3900 K < T_{eff} \leq 7300 K \\
        0.553\left[\left(\frac{a}{a_\oplus}\right)^{1.003}\left(\frac{R_p}{R_\oplus}\right)^{-0.894}\right]g_1\left(t\right) & R_p \geq 2.771R_\oplus, \quad 3900 K < T_{eff} \leq 7300 K
    \end{cases}
\end{equation}
where
\begin{equation}
    \begin{split}
        g_0\left(t\right) &= 1 - 3.572t - 1.327t^2 \\
        g_1\left(t\right) &= 1 - 2.538t - 1.660t^2
    \end{split}
\end{equation}
and the numerical values are the 50th percentile values from the SMM and BRMt2FGK models. There is strong evidence that M type stars have more planets on short orbital periods \citep{mulders2015increase,burke2015terrestrial} than F, G, and K type stars, so it is reasonable that occurrence rate density at 3900 K is piecewise continuous. Our analytical best fit model has the benefit that simple analytical conditional probability densities may be formed which allow planet samples to be generated quickly by Gibbs sampling \citep{gelfand1990sampling} with self-written code or existing packages like \texttt{BUGS} \citep{lunn2009bugs} or \texttt{JAGS} \citep{plummer2003jags}. Sampling from published occurrence rates as in \cite{barclay2018revised} requires an assumption of how planets are distributed within each bin (e.g. uniform or log-uniform), whereas sampling from our analytical best-fit model requires no such assumption.

\section{Discussion}\label{sec:discussion}
We now compare our model fits to literature data. \cite{kopparapu2018exoplanet} noted that combinations of SAG13 occurrence rate data tend to fall between \cite{petigura2013prevalence} at the low end and \cite{burke2015terrestrial} at the high end. The low end occurrence rate calculations from \cite{petigura2013prevalence} were based on an early and incomplete planet catalog. The high end occurrence rate calculations from \cite{burke2015terrestrial} considered many additional factors affecting occurrence rates and are more robust. We included the SAG13 data from Burke in our model fits, so we anticipate that our results derived from our model fits, will be close to those of \cite{burke2015terrestrial}.

We first compare our fit results to \cite{dressing2015occurrence} occurrence rates for M stars to highlight the break in the $ T_{eff} $ trend for stars with lower $ T_{eff} $. \cite{dressing2015occurrence} updated the results of \cite{dressing2013occurrence} and gave slightly higher values for the occurrence rates. A comparison of our model fits to \cite{dressing2015occurrence} is shown in Table~\ref{tab:dressing}. We report the occurrence rates calculated at the 50th percentile model parameter values and include the 16th and 84th percentile values as our uncertainty estimate. The SMtAll model underpredicts the \cite{dressing2015occurrence} occurrence rates by a factor between three and seven. The BRMtAll model performs better than the SMtAll model, but still underpredicts the \cite{dressing2015occurrence} occurrence rates by a factor of two to five. The uncertainty levels of the SMM model overlap with those from \cite{dressing2015occurrence} to give our best fit to the M type star data. This shows that the dependence on $ T_{eff} $ we assumed must have a break to accommodate M type stars with lower $ T_{eff} $.

\afterpage{
\begin{landscape}
\begin{table}
    \caption{Occurrence rate comparison for M type stars. The uncertainty bounds on our model results come from the 16th and 84th percentile model parameter values.}
    \centering
    \begin{tabular}{cccccc}
        \hline\hline
        & & Dressing and & & & \\ 
        & & Charbonneau~\cite{dressing2015occurrence} & SM$\tau$All & BRM$\tau$All & SMM  \\
        \hline
        $ P \in \left[0.5,50\right] $ days & $ R_p \in \left[1,1.5\right]R_\oplus $ & $ 0.56_{-0.05}^{+0.06} $ & $ 0.133_{-0.012}^{+0.012} $ & $ 0.154_{-0.017}^{+0.020} $ & $ 0.486_{-0.171}^{+0.253} $ \\
        $ P \in \left[0.5,50\right] $ days & $ R_p \in \left[1.5,2\right]R_\oplus $ & $ 0.46_{-0.05}^{+0.07} $ & $ 0.062_{-0.006}^{+0.007} $ & $ 0.103_{-0.013}^{+0.016} $ & $ 0.277_{-0.107}^{+0.168} $ \\
        $ P \in \left[0.5,200\right] $ days & $ R_p \in \left[1,4\right]R_\oplus $ & $ 2.5_{-0.2}^{+0.2} $ & $ 0.837_{-0.072}^{+0.077} $ & $ 1.107_{-0.159}^{+0.162} $ & $ 4.095_{-1.427}^{+2.131} $ \\
        \hline
    \end{tabular}
    \label{tab:dressing}
\end{table}
\end{landscape}
}

We compare our models to habitable zone occurrence rates reported by SAG13 and \cite{burke2015terrestrial}. The habitable zone was calculated for a solar twin ($ t = 0 $) from \cite{kopparapu2013habitable}. These occurrence rates are given for a conservative (338--792 days or 0.95--1.68 AU) and an optimistic (237--864 days or 0.75--1.78 AU) estimate of the habitable zone. This comparison is shown in Table~\ref{tab:habitable}. Our BRMt2FGK results are higher than the SAG13 and \cite{burke2015terrestrial} results but are still within the uncertainty ranges, thus showing agreement with these results.

\afterpage{
\begin{landscape}
\begin{table}
    \caption{Occurrence rate comparison to SAG13 and \cite{burke2015terrestrial} for estimates of the habitable zone for a G dwarf star. The conservative estimate covers a range of 338--792 days or 0.95--1.68 AU and the optimistic estimate covers a range of 237--864 days or 0.75--1.78 AU. The habitable zone boundaries were calculated from \cite{kopparapu2013habitable}. The uncertainty bounds on our BRM$\tau$2FGK model results come from the 16th and 84th percentile model parameter values.}
    \centering
    \begin{tabular}{ccccc}
        \hline\hline
        & & SAG13 & \cite{burke2015terrestrial} & BRM$\tau$2FGK \\
        \hline
        \multirow{2}{*}{Conservative} & $ R_p \in \left[1,1.5\right]R_\oplus $ & $ 0.14_{-0.04}^{+0.12} $ & $ 0.21_{-0.08}^{+0.08} $ & $ 0.31_{-0.03}^{+0.02} $ \\
        & $ R_p \in \left[0.5,1.5\right]R_\oplus $ & $ 0.40_{-0.14}^{+0.48} $ & $ 0.50_{-0.20}^{+0.40} $ & $ 0.88_{-0.03}^{+0.04} $ \\
        \hline
        \multirow{2}{*}{Optimistic} & $ R_p \in \left[1,1.5\right]R_\oplus $ & $ 0.20_{-0.06}^{+0.18} $ & $ 0.31_{-0.10}^{+0.10} $ & $ 0.43_{-0.03}^{+0.03} $ \\
        & $ R_p \in \left[0.5,1.5\right]R_\oplus $ & $ 0.58_{-0.20}^{+0.70} $ & $ 0.73_{-0.30}^{+0.60} $ & $ 1.24_{-0.05}^{+0.06} $ \\
        \hline
    \end{tabular}
    \label{tab:habitable}
\end{table}
\end{landscape}
}

\begin{figure}[ht]
\figurenum{13}
\label{fig:burkecomp}
    \plotone{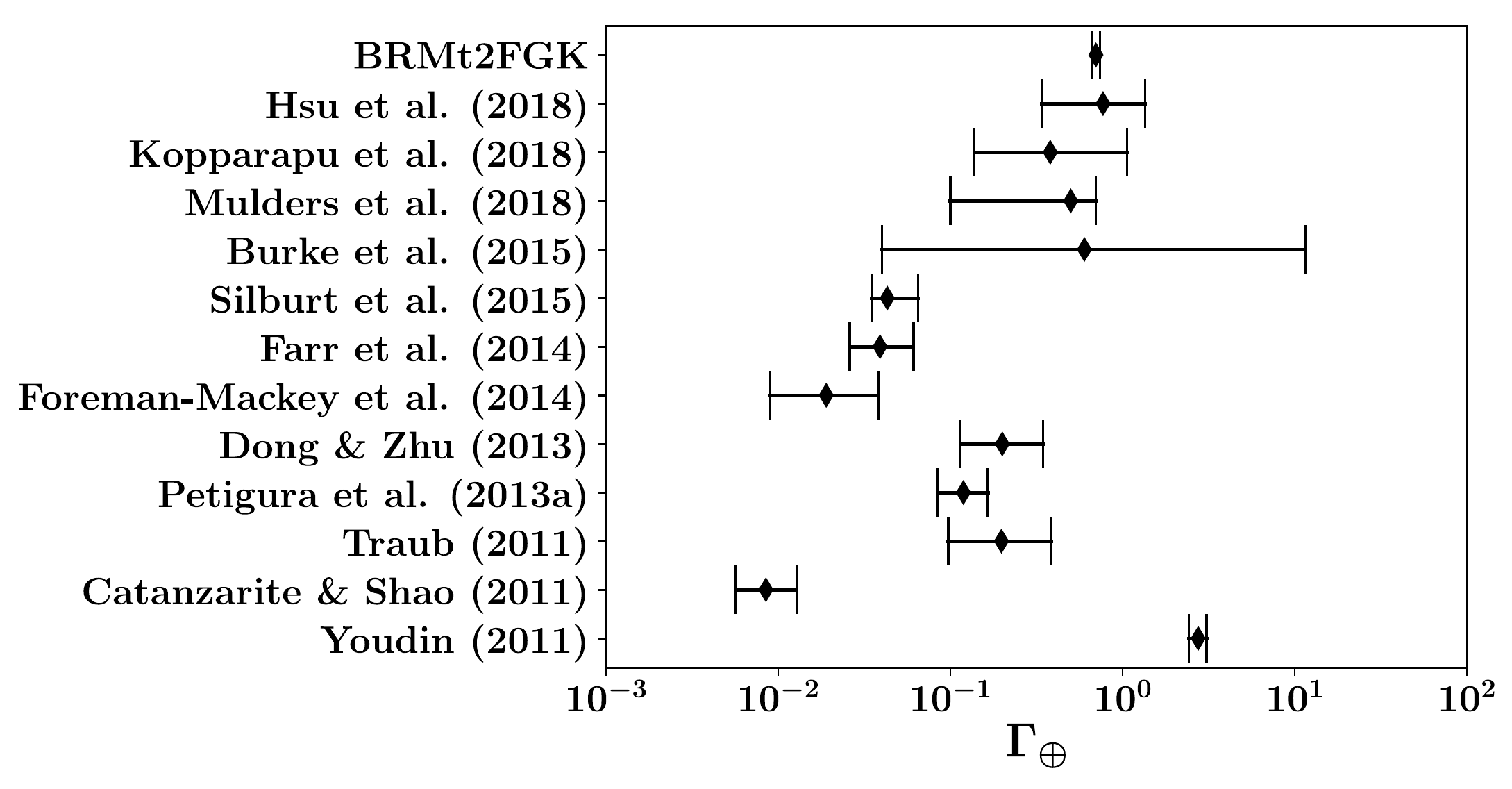}
    \caption{Comparison of $ \Gamma_\oplus $ from the BRMt2FGK model and values from the literature. This figure is similar to figures found in \cite{foreman2014exoplanet} and 
    one generated by Leslie Rogers (SAG13 close-out presentation)}
\end{figure}

Our final comparison is to $ \Gamma_\oplus $ \citep{youdin2011exoplanet,foreman2014exoplanet,burke2015terrestrial} results from a number of studies where
\begin{equation}
    \Gamma_{\oplus} = \left.\frac{\partial^2 \eta}{\partial \ln P \partial \ln R_p}\right\vert_{1\;\mathrm{year},1\;R_\oplus}.
\end{equation}
We perform a change of variables to get our models in the proper form
\begin{equation}
    \frac{\partial^2 \eta}{\partial \ln P \partial \ln R_p} = \frac{2C_i}{3}\left(\sqrt[3]{\frac{GM}{4\pi^2}}\right)^{\alpha_i}\left(\frac{P}{P_\oplus}\right)^{\frac{2\alpha_i}{3}}\left(\frac{R_p}{R_\oplus}\right)^{\beta_i} g\left(t\right) \,.
\end{equation}
For sun-like stars, this evaluates to
\begin{equation}\label{eq:gammaE}
    \Gamma_\oplus = \frac{2C_0}{3}.
\end{equation}
Figure~\ref{fig:burkecomp} shows a comparison of our results to the literature and is similar to figures found in \cite{foreman2014exoplanet} and one generated by Leslie Rogers (included in the SAG13 close-out presentation). We present $ \Gamma_\oplus $ results for our BRMt2FGK model. The error bars for our model are based on the 16\% and 84\% values of $ \ln C $. The values from \cite{hsu2018improving} when scaled to the appropriate units are $ 0.77_{-0.43}^{+0.58} $. The \cite{kopparapu2018exoplanet} values are the reported $ 0.38_{-0.242}^{+0.68} $ which represent the SAG13 baseline, pessimistic, and optimistic cases. The values from \cite{mulders2018exoplanet} reflect the 1$ \sigma $ values from the fitted parameters. The values from \cite{burke2015terrestrial} represent the allowable range. The \cite{foreman2014exoplanet} values are the reported $ 0.019_{-0.010}^{+0.019} $ values. The \cite{dong2013fast} values come from an extrapolation of the 1--2$ R_\oplus $ fits where the error bars are given by the 1$ \sigma $ values reported in their Table 2 and scaled to the appropriate units. The \cite{petigura2013prevalence} values reflect the reported $ 0.119_{-0.035}^{+0.046} $. The \cite{traub2011terrestrial} values come from an extrapolation of Equation 2 (including the factor $ \rho_\oplus \simeq 0.291 $) and the error bars come from \cite{traub2011terrestrial} Equation 6 and Equation 7. The \cite{catanzarite2011occurrence} values come from an extrapolation of the power law fit to period and the error bars reflect the 1$ \sigma $ values on the power law index. The values from \cite{youdin2011exoplanet} are the reported $ 2.75_{-0.33}^{+0.33} $. 

We see that $ \Gamma_\oplus $ from the BRMt2FGK model compares favorably with the values from \cite{hsu2018improving}, \cite{kopparapu2018exoplanet}, \cite{mulders2018exoplanet}, and \cite{burke2015terrestrial} and is well within the allowable range. All of the $ \Gamma_\oplus $ ranges from the literature fall within the allowable range of \cite{burke2015terrestrial} except for the $ \Gamma_\oplus $ ranges from \cite{foreman2014exoplanet} and \cite{catanzarite2011occurrence}. \cite{burke2015terrestrial} noted that there is overlap in the upper tail of $ \Gamma_\oplus $ from \cite{foreman2014exoplanet} to the allowable range. \cite{foreman2014exoplanet} used the same inputs as \cite{petigura2013prevalence} but found a steeper fall off of occurrence rates at longer periods. \cite{foreman2014exoplanet} also found that accounting for uncertainty on planet radii led to a systematically lower $ \Gamma_\oplus $ value than \cite{petigura2013prevalence}. \cite{burke2015terrestrial} noted that further work is needed to determine whether the differences between the \cite{foreman2014exoplanet} result and the others from the literature come from differing inputs or methodology. \cite{catanzarite2011occurrence} fit a power law model on orbital period for occurrence rate data for 2--4$ R_\oplus $ planets with periods up to 132 days. The $ \Gamma_\oplus $ values reported here come from their single power law fit and extrapolated to a planetary radius (unaccounted for in their power law model on period) of $ R_\oplus $ and period of one year. Because we extrapolate both planetary radius and period to get $ \Gamma_\oplus $, it is understandable that \cite{catanzarite2011occurrence} is an outlier. 

Our result shows the smallest error bars on $ \Gamma_\oplus $ because the only source of uncertainty from our model fits reduced to Equation~\ref{eq:gammaE} is in the $ C_0 $ term. The set up of our models was specifically chosen (transforming $ T_{eff} $ values to $ t $ where $ T_{eff} = T_{eff,\odot} $ results in $ t = 0 $) to be most accurate for sun-like stars. When including the 16th and 84th percentile values for the BRMt2FGK model $ \ln C_0 = 0.049_{-0.056}^{+0.056} $, $ C_0 = 1.050_{-0.057}^{+0.060} $, and $ \Gamma_\oplus = 0.700_{-0.038}^{+0.040} $. Although we sampled the posterior distribution of $ \ln C_0 $, these values show that propagating $ \ln C_0 $ to $ \Gamma_\oplus $ results in narrow 1$ \sigma $ error bars.

\section{Conclusions}
We have presented models for planet occurrence rate density based on previous occurrence rate calculations consisting of power laws on semi-major axis and planetary radius and a polynomial function of stellar $ T_{eff} $. We found that M type stars do not follow the same relation on $ T_{eff} $ as F, G, and K type stars. We found that the best model fit to the M type star data was the SMM model, a power law on semi-major axis and planetary radius. We found that the best model fit to the F, G, and K type star data was the BRMt2FGK model, power laws broken at 2.771$ R_\oplus $ with a quadratic function of $ T_{eff} $. Our models give occurrence rates that are comparable with other published results and explicitly include stellar effective temperature as a variable. By explicitly including stellar effective temperature in our models, we are able to fit a wider range of occurrence rate data than previously published models. 

By including more than just planetary physical and orbital parameters, our models are a step toward a more complete model of planet occurrence rates. The next step will be to create a model including additional stellar parameters which affect occurrence rates. Using our models in mission simulation or science yield calculations will give more accurate occurrence rates for individual stars by considering their individual stellar effective temperatures in the generation of planet samples instead of treating all stars the same. This will lead to more accurate science yield estimations for future and proposed exoplanet finding missions like HabEx and LUVOIR.

\acknowledgments
The authors would like to thank the anonymous reviewer for their constructive comments. We thank Natalie Batalha, Chris Burke, and Gijs Mulders for the use of their SAG13 data. We also gratefully acknowledge all who contributed to the SAG13 community occurrence rate project.

%\bibliography{occbib}
%\bibliographystyle{aasjournal.bst}

\end{document}